\documentclass[aps,prb,reprint,showpacs]{revtex4-1}

\usepackage[utf8]{inputenc}
\usepackage[english]{babel}
\usepackage{amssymb,amsmath,mathrsfs}
\usepackage{textcomp}
\usepackage{bm}
\usepackage{graphicx}
\usepackage{color}

\begin{document}

\title{Dynamic magnetostriction for antiferromagnets}

\author{Thomas Nussle}
\email{thomas.nussle@cea.fr}
\affiliation{CEA DAM/Le Ripault, BP 16, F-37260, Monts, FRANCE}
\affiliation{Institut Denis Poisson, Université de Tours, Université d'Orléans, CNRS (UMR7013), Parc de Grandmont, F-37200, Tours, FRANCE}
\author{Pascal Thibaudeau}
\email{pascal.thibaudeau@cea.fr}
\affiliation{CEA DAM/Le Ripault, BP 16, F-37260, Monts, FRANCE}
\author{Stam Nicolis}
\email{stam.nicolis@lmpt.univ-tours.fr}
\affiliation{Institut Denis Poisson, Université de Tours, Université d'Orléans, CNRS (UMR7013), Parc de Grandmont, F-37200, Tours, FRANCE}

\begin{abstract}
In this paper we study the switching properties of the dynamics of magnetic moments, that interact with an elastic medium. 
To do so we construct a Hamiltonian framework, that can take into account the dynamics in phase space of the variables that describe the magnetic moments in a consistent way. 
It is convenient to describe the magnetic moments as bilinears of anticommuting variables that are their own conjugates. 
However, we show how it is possible to avoid having to deal directly with the anticommuting variables themselves, only using them to deduce non-trivial constraints on the magnetoelastic couplings. 
We construct the appropriate Poisson bracket and a geometric integration scheme, that is symplectic in the extended phase space and that allows us to study the switching properties of the magnetization, that are relevant for applications, for the case of a toy model for antiferromagnetic NiO, under external stresses.

In the absence of magnetoelastic coupling, we recover the results reported in the literature and in our previous studies. 
In the presence of the magnetoelastic coupling, the characteristic oscillations of the mechanical system have repercussions on the Néel order parameter dynamics. 
This is particularly striking for the spin accumulation which is more than doubled by the coupling to the strain ; here as well, the mechanical oscillations are reflected on the magnetic dynamics. 
As a consequence of such a stress induced strain, the switching time of the magnetization is slightly faster and the amplitude of the magnetization enhanced. 

\end{abstract}

\maketitle

\section{Introduction\label{introduction}}

With the emergence of spintronics, attention has now been focused on the theoretical understanding and experimental manipulation of the properties of magnetic materials at ever shorter length and time scales. 
In this context, antiferromagnets (AFs) are the magnetically ordered materials which appear to be among the most promising candidates for realizing fast spintronic devices with very low power consumption~\cite{raduTransientFerromagneticlikeState2011,*jungwirthAntiferromagneticSpintronics2016}. 
Therefore quantitative modeling of spin transport is also of topical interest. 

Recent progress in describing spin transport and spin-transfert torque (STT) effects in such AF devices opens a route towards developing multilevel memory devices with switching speeds, that could exceed those of devices made of ferromagnetic materials and semiconductors~\cite{zeleznySpinTransportSpin2018}.
Moreover reversible strain-induced magnetization switching in ferromagnetic materials has been reported that also allows the design of a rewritable, non-volatile, non-toggle and extremely low energy straintronic memory~\cite{ahmadReversibleStraininducedMagnetization2015}.
Recently a piezolectric, strain-controlled AF memory, insensitive to magnetic fields, has been tested, as an example of controlling magnetism by electric fields in multiferroic heterostructures~\cite{yanPiezoelectricStraincontrolledAntiferromagnetic2019}.

However a theoretical description of all these phenomena is, still, very much, work in progress, since, at the scales that are relevant, it is not possible to separate the magnetic, electric and elastic responses--all must be treated simultaneously.

It is, indeed, noteworthy that no direct local coupling between electromagnetic field components is allowed in vacuum, at the classical level; the magnetic response to an electric field is necessarily mediated by atomic position arrangements of magnetic moments, which in response produce non-local mechanical strains~\cite{fiebigRevivalMagnetoelectricEffect2005,atulasimhaBennettClockingNanomagnetic2010}.  
This response is known, more generally, under the heading of ``magneto--elasticity''~\cite{dutremoletdelacheisserieMagnetostrictionTheoryApplications1993}.

Magneto-elastic phenomena have been, typically, viewed either from the perspective of continuum mechanics, where the magnetic properties of materials are incorporated into constitutive nonlinear laws of electro-conductive bodies~\cite{mauginContinuumMechanicsElectromagnetic1988}, or through the introduction of effective magnetic anisotropies mimicking the static strains of a solid~\cite{kittelPhysicalTheoryFerromagnetic1949}.

Most studies, however, do not investigate the backreaction of the magnetization on the mechanical strains; however it is such a backreaction that can become significant at very short length and time scales (where a continuum description, anyway reaches its limits). 
This highlights the need for developing multiscale models. 

Even if multiscale models have been investigated for magneto-elastic couplings~\cite{buironMultiscaleModelMagnetoelastic1999}, with good experimental comparison for single crystals and polycrystalline samples~\cite{danielConstitutiveLawMagnetostrictive2008}, insights from the framework of localization and homogenization are still the mainstays for deducing the constitutive laws for polycrystalline media.
Moreover, in this context, homogenization methods for multiscale mechanics assume the existence of well--separated scales and different relations among length scales, that lead to different effective equations, which, in turn, represent the corresponding different physical effects, appropriate at each scale~\cite{meiHomogenizationMethodsMultiscale2010}.

Renormalization group ideas, of course, are instrumental in providing a description of how the dynamics depends on the scale; what has been lacking, to date, is a unified construction of the dynamics of elastic and spin degrees of freedom, within a common Hamiltonian approach, at any particular scale. 
In particular, what has, curiously, not been fully exploited is the resolution of the constraints, that the spin degrees of freedom are subject to, since they are their own canonically conjugate variables, as their equations of motion are first order.
This means that they can be most usefully expressed as multilinear combinations of {\em anticommuting} variables. 
As we shall see, this can lead to useful insights, in practice, about the dynamics of the coupled system. 

While the generalization of the canonical formalism, that incorporates the constraints implied by spin degrees of freedom has been worked out a long time ago~\cite{casalbuoniQuantizationSystemsAnticommuting1976,*casalbuoniClassicalMechanicsBosefermi1976,
*berezinParticleSpinDynamics1977,*nelsonLagrangianTreatmentMagnetic1994}, it has not been applied to concrete situations of physical interest, simply because these, up to now, were not of practical relevance. 

In the present paper we provide a complete description of the combined Hamiltonian dynamics of the elastic and of the spin degrees of freedom in phase space and show how the anticommuting nature of the spin degrees of freedom can be--indirectly--tested. 
It is the refined control over these and the dynamical interplay between strain and magnetization that are the particular novelties of our approach. 

The elastic (local mechanical strain) and spin degrees of freedom are thus assigned to sites of a lattice. 
We deduce the corresponding fully coupled equations of motion from the corresponding Hamiltonian formalism, that treats mechanical and spin degrees of freedom in a unified way and study the dynamics of this system, comparing it to previous studies, first for the purely magnetic part~\cite{chengUltrafastSwitchingAntiferromagnets2015} and then, also, for the magneto-elastically coupled system~\cite{nussleCouplingMagnetoelasticLagrangians2019}. 

Of course to render the equations tractable we must resort to approximations.
The approximation we employ is the mean field approximation, that amounts, in the present case, to a coarse--graining of the lattice, which means that the sites are, in fact, domains, within each of which the properties are homogeneous. 
Within this framework, our Hamiltonian formalism is, however, exact. 

The paper is organized as follows: In section~\ref{section1} we construct a classical Hamiltonian for the mechanical, and the spin degrees of freedom and discuss their couplings. We discuss the advantages of describing the magnetic moments in terms of anticommuting, Majorana,  variables, whose salient properties are reviewed in appendix~\ref{appendix1}. 

In section~\ref{section2}, we introduce a Poisson bracket, which allows us to obtain the equations of motion for the time-evolution of all 
the dynamical variables.
These are, then solved, in section~\ref{section3} using a symplectic integration scheme. 
Special attention is devoted to the stability analysis of the schemes and to the order of the different symplectic operators. 
This analysis is presented in appendix~\ref{appendix2}. 

Finally in section~\ref{section4} we illustrate the formalism for the case of a simple toy model for NiO, consisting of two spins, interacting through antiferromagnetic exchange, and switched by an external STT, as well as under an external stress. 
It is here, in particular, that we show that the formalism is capable of handling the interaction between two domains, taking into account the coupling between elastic and magnetic degrees of freedom. 

\section{The Hamiltonian\label{section1}}

In this section we propose a Hamiltonian for the combined system of elastic and spin degrees of freedom. 
These parts have, of course, been considered before~\cite{dutremoletdelacheisserieMagnetostrictionTheoryApplications1993}, so we shall review the salient features, before introducing our approach.

The starting point for the elastic degrees of freedom is the framework of mechanics for elastic solids, within the regime of the approximation of small deformations, supplemented by the assumption of perfect mechanical micro-reversibility~\cite{casimirOnsagerPrincipleMicroscopic1945,*landauTheoryElasticity2008}.
These statements imply that an elastic material is characterized by a collection of $N$ time-dependant elastic deformations, which are described by a set of symmetric Cauchy strain tensors ${\epsilon_{IJ}^{i}(t)}$ where $I,J$ are spatial indices, ranging from 1 to 3, and assigned to each lattice site, $i\in \{1..N\}$, of the material.
At equilibrium, each of these variables is independent of time and we assume that such a state not only exists, but can be relevant for the time scales of interest~\cite{mauginContinuumMechanicsElectromagnetic1988}. 

The total internal energy of this system can be described in terms of the  interactions between these variables, that take into account the elastic mechanical part in the most general form and the interaction with external stresses, which will be described by the magnetic degrees of freedom.  Schematically,
\begin{equation}
\label{Htot}
\frac{\mathcal{H}}{V_0}=\frac{\mathcal{H}_\mathrm{mech}}{V_0}+\frac{\mathcal{H}_\mathrm{ext}}{V_0}
\end{equation}
We shall now spell out each term. 

To lowest, non--trivial, i.e. quadratic, order, $\mathcal{H}_\mathrm{ext}/V_0$ can be written as~\cite{gairolaNonlocalTheoryElastic1978,*eringenNonlocalContinuumField2004}
\begin{equation}
	\label{mechanicenergyfull}
	\frac{{\cal{H}}_\textrm{mech}}{V_0}=\frac{1}{2}\sum_{i =1}^NC_{IJKL}^{i}\epsilon_{IJ}^i(t)\epsilon_{KL}^i(t),
\end{equation}
where an implicit sum on the repeated space indices is understood and where $C_{IJKL}^{i}$ are the elastic constants with $V_0$ as the reference volume.

In this expression we have assumed that the different sites of the material do not interact through elastic deformations (they will only interact through spin exchange forces, to be spelled out below).

For an homogeneous material ({\it i.e.}, $C^{(i)}=C$ for every site $i$), these elastic constants can be expressed in terms of the two Lamé parameters~\cite{landauTheoryElasticity2008}, designated by $(C_0,C_1)$, which in Cartesian coordinates read as follows:
\begin{equation}
\label{lame}
C_{IJKL}=C_0\delta_{IJ}\delta_{KL}+C_1\left(\delta_{IK}\delta_{JL}+\delta_{IL}\delta_{JK}\right).
\end{equation}
This mechanical system can be excited by an external stress $\sigma_{IJ}^\textrm{ext}$, through the coupling 
\begin{equation}
\label{stress}
\frac{{\cal H}_\textrm{ext}}{V_0}=-\sum_{i=1}^N\sigma_{IJ}^\textrm{ext}\epsilon_{JI}^i.
\end{equation}
While this stress can be, in general, a time- and space-dependent quantity, it is, nonetheless, assumed to be uniform over the sites of the material. 
 
The magnetic degrees of freedom, noted $S_I^i(t)$, are assigned to  the same lattice sites as the strain and describe how their dynamics affects the mechanical degrees of freedom, through the so-called magnetoelastic coupling.

The contribution of the magnetoelastic coupling to the Hamiltonian can be deduced by assuming (global) Galilean invariance on the one hand and imposing invariance under the symmetries of the point group~\cite{toupinSoundWavesDeformed1961,*callenStaticMagnetoelasticCoupling1963,*brownMagnetoelasticInteractions1966} on the other. 
For an isotropic medium, these requirements lead to the expressions for  the total energy and the total magnetization~\cite{rosenGalileanInvarianceGeneral1972}.
The total internal energy is, thus, a sum over the lattice~\cite{tierstenCoupledMagnetomechanicalEquations1964,*akhiezerSpinWaves1968}.

Consequently, in the approximation of small deformations, the expansion of the magnetoelastic coupling energy will contain even powers of the magnetization ${\bm S}^{i}$. 
It will, however, contain all powers of the strain tensor, $\epsilon$. 

The linear part in $\epsilon_{IJ}^i$ defines the first-order, magnetoelastic coupling, which is responsible for the magnetostriction, while the quadratic terms in $\epsilon_{IJ}^i$ define the second-order magnetoelastic coupling, that is responsible for the ``nonlinear response'' in the elastic properties of magnetic media, that have been, also, identified as describing ``morphic effects''~\cite{dutremoletdelacheisserieMagnetostrictionTheoryApplications1993}. 

Higher order terms are, usually, ignored, since their coefficients are assumed to be much smaller than the lower order terms, already discussed. 

Therefore, the contribution to the--total--Hamiltonian of the magnetoelastic coupling can be written as
\begin{align}
	\label{magnetoelasticenergy}
	{\cal H}_{\textrm{ME}}&=\sum_{i,j=1}^N B_{IJKL}^{(1)i,j}\epsilon_{IJ}^i(t)S_K^i(t)S_L^j(t)\nonumber\\
	&+B_{IJKLMN}^{(2)i,j}\epsilon_{IJ}^i(t)\epsilon_{KL}^j(t)S_M^i(t)S_N^j(t)+\dots
\end{align}
The total Hamiltonian can, thus, be written as 
\begin{equation}
\label{Htot1}
\mathcal{H}_\mathrm{tot} = \mathcal{H}_\mathrm{mech}+\mathcal{H}_\mathrm{ext}+\mathcal{H}_\mathrm{ME}
\end{equation}
and leads to a ``renormalization'' of the effective elastic constants in the following way: If one compares the last term of equation \eqref{magnetoelasticenergy} with equation \eqref{mechanicenergyfull}, it is easy to see that a pair of effective elastic constants can be introduced that depends on the value of the spins on each site, i.e.
\begin{equation}
	C_{IJLK}^{\textrm{eff},i,j}\equiv C_{IJLK}^{ij}+\frac{1}{V_0}B_{IJKLMN}^{(2)i,j}S_M^i(t)S_N^j(t).
\end{equation}
This has been used to provide an interpretation of the ``anomalous'' temperature dependence in the elastic constants in iron single crystals, for instance, as resulting from the competition between spin ordering and diffusion effects~\cite{deverTemperatureDependenceElastic1972}. 
In the present work however, we will only consider $B_{IJKL}^{(1)i,j}$ and set aside the non-linear mechanical terms. 
Although there is no particular difficulty to take the latter into account in this formalism, we will take $B^{(2)}\equiv 0$ for the rest of this study and for the sake of simplicity.

In Appendix {\ref{appendix1}} we recall that the most effective ``classically  equivalent description'' of the spin degrees of freedom, that can capture consistently their interaction with the elastic degrees of freedom is not through the variables ${\bm S}$ but through their ``Doppelgänger'' ${\bm\xi}$, related to the $S_I^i$ through 
\begin{equation}
\label{Sxip}
S_I^i\equiv-\frac{\imath}{2}\epsilon_{IJK}\xi_J^i\xi_K^i.
\end{equation}
The $\bm{\xi}$ can be identified with Majorana fermions, which have found many applications recently in condensed matter systems~\cite{wilczekMajoranaReturns2009,*aliceaNewDirectionsPursuit2012}, where new methods for controlling spin degrees of freedom have been developed.

It is interesting to remark that the anticommuting variables, $\xi_k^i$ are not Grassmann variables, satisfying $\{\xi_I^i,\xi_J^j\}=0$; but, rather, 
$\{\xi_I^i,\xi_J^i\}=\delta^{ij}\delta_{IJ}$, i.e. that the $\xi_I$ generate a Clifford algebra on each site~\cite{lounestoCliffordAlgebrasSpinors2003}.
It is in this way that the $S_I^i$, defined through eq.~(\ref{Sxip}), satisfy the angular momentum algebra~\cite{varshalovichQuantumTheoryAngular1988}.   
 
If one were tempted to simply replace ${\bm S}$ by ${\bm \xi}$ in eq.\eqref{magnetoelasticenergy}, it is interesting to remark that, for $N=1$, and because of the symmetries of $B^{(1)}$ as recalled in reference~\cite{dutremoletdelacheisserieMagnetostrictionTheoryApplications1993}, ${\cal H}_{ME}=0$, which implies that the dynamics as it is cannot be encoded by this Hamiltonian.

On the other hand, this allows us to understand the constraints on the allowed terms in the true Hamiltonian. 
They must, necessarily, involve more than one spins. 
Indeed, we can construct expressions that are multi-linear combinations of the $S^i_I$ on different sites, potentially, up to order $N$, the number of sites, since no two identical $\xi_I^i$ variables occur in the same monomial. 
This is, therefore, a nice way of automatically organizing the multi-spin terms of the Hamiltonian. 

In the formalism of Atomistic Spin Dynamics, since the magnetic moments are localized~\cite{evansAtomisticSpinModel2014,erikssonAtomisticSpinDynamics2017}, it is customary to consider spatial averages, around each site, defining an effective macroscopic localized spin
\begin{equation}
\label{Sxi}
 S_I^{\textrm{eff}}=\langle S_I^i\rangle_i
\end{equation}
which implies that multi-linear expressions, $S_IS_J$ no longer vanish identically, as was imposed by the anticommuting nature of $\bm{\xi}$ before. We can, thus, understand the relevance of this averaging procedure, in terms of the description of the spin degrees of freedom in terms of the anticommuting variables. 

The justification for equation~\eqref{magnetoelasticenergy} has been advocated a long time ago~\cite{vanvleckAnisotropyCubicFerromagnetic1937,*neelAnisotropieMagnetiqueSuperficielle1954,*leeMagnetostrictionMagnetomechanicalEffects1955}, as stemming from a model for a two--body interaction, that is itself a pedagogical version of the quantum theory of interacting magnons and phonons~\cite{sabiryanovMagnonsMagnonPhononInteractions1999}. 
In eq.~(\ref{magnetoelasticenergy}) what has been left unspecified are the properties of the tensor $B^{(1)}$ under exchange of the indices $i$ and $j$, that label the sites. 
What is customary is to assume that $B^{(1)}$, in fact, does not depend on $i$ and $j$ at all--it is homogeneous across the material. 
By stopping the expansion at first order in the mechanical deformation in~\eqref{magnetoelasticenergy} and assuming homogeneity in the material, we, therefore, end up with the following expression
\begin{equation}
	{\cal H}_{\textrm{ME}}=B_{IJKL}^{(1)}\epsilon_{IJ}(t)S_K(t)S_L(t)\label{MEfirstOrder}.
\end{equation}
Again, in the case of an isotropic medium, the elements in $B^{(1)}$ enjoy the same symmetries than the elastic constants and can be expressed only in terms of two constants $B^{(1)}_0$ and $B^{(1)}_1$ by the following expression
\begin{equation}
	B_{IJKL}^{(1)}=B^{(1)}_0\delta_{IJ}\delta_{KL}+B^{(1)}_1\left(\delta_{IK}\delta_{JL}+\delta_{IL}\delta_{JK}\right).
\end{equation}
In summary, the microscopic theory underlying Eq.\eqref{MEfirstOrder} describes the interaction of three particles, two of which describe  spin states, and one  the state of the elastic deformation. 
However the spin states, are, in fact, bound states of more ``fundamental'' entities, the anticommuting $\bm{\xi}$. 
$B^{(1)}$ can then be interpreted as the vertex of this interaction, that describes a spinning particle, that does not directly interact with itself--since any such interaction is inconsistent with the fact that the $\bm{\xi}$ anticommute. 
Such a self-interaction can, however, appear on larger scales, when spatial averages can become meaningful for describing the spin degrees of freedom. 

In conclusion to this section, the Hamiltonian we have constructed describes the interaction of magnetic moments through their embedding within an elastic medium. 
What remains to be done is to define the Poisson brackets, that can take into account the evolution in phase space of commuting, as well as anticommuting degrees of freedom. 
Indeed, the construction procedure for commuting canonical variables is well known, however, including anticommuting variables in a unified way has remained a rather esoteric subject--known in theory~\cite{casalbuoniQuantizationSystemsAnticommuting1976,casalbuoniClassicalMechanicsBosefermi1976,
berezinParticleSpinDynamics1977}, not, however, implemented, in practice.

In the following section, we shall construct the equations of motion in the phase space of the elastic and the spinning degrees of freedom, that 
implements these ideas in practice. 

In order to work directly with the anticommuting variables themselves, in combination with the mechanical degrees of freedom requires implementing  a ``graded Poisson bracket''; one way to do this is discussed in Appendix~\ref{appendix1}.

\section{The Poisson brackets and the equations of motion\label{section2}}

To deduce from the Hamiltonian, discussed in the previous section, the equations of motion, we must define the appropriate pairs of canonically conjugate variables and consequently their Poisson bracket.

First we recognize that $\varepsilon$ acts as a tensor and one can understand the definition of the Poisson bracket of rank-2 symmetric tensors as an application of the DeDonder-Weyl covariant hamiltonian formulation of field theory~\cite{kastrupCanonicalTheoriesLagrangian1983,*kanatchikovCanonicalStructureDonderWeyl1993}.
Although the context is different, the ADM procedure in general relativity also provides such a Poisson bracket~\cite{arnowittCanonicalVariablesGeneral1960} (with further relations, between the conjugate variables that are not relevant here).

Although no clear consensus has, in fact, emerged on the properties of Poisson bracket of rank-2 tensors~\cite{hojmanGeometrodynamicsRegained1976}, 
if one focuses on the special case of strain tensors, that depend only on time, the following conjugate momentum can be introduced
\begin{equation}
	\label{canonicalpi}
	\pi_{IJ}\equiv\frac{\partial\mathscr{L}}{\partial{\dot{\varepsilon}_{IJ}}},
\end{equation}
where $\mathscr{ L}(\varepsilon_{IJ}(t),\dot{\varepsilon}_{IJ}(t))$ is the unconstrained and free Lagrangian.
$\dot{\varepsilon}_{IJ}(t)$ are the components of the strain-rate tensor~\cite{hillConstitutiveInequalitiesIsotropic1970}.  

Thus, we can build the corresponding Hamiltonian ${\cal H}$ for the time evolution with tensor variables for mechanical deformations $\bm{\varepsilon}$ and their conjugated momenta $\bm{\pi}$, because these quantities admit unbounded numerical values, as the corresponding Legendre transform 
\begin{equation}
	\label{legendretransform}
	{\cal H}({\bm\varepsilon},{\bm\pi})=\pi_{IJ}\dot{\varepsilon}_{IJ}-\mathscr{ L},
\end{equation}
up to a total time derivative for ${\mathscr L}$.

For the mechanical system only (i.e. for functions $A({\bm\varepsilon},{\bm\pi})$ and $B({\bm\varepsilon},{\bm\pi})$), the Poisson bracket can be defined in perfect analogy to that of any particle system, that explores a given target space geometry (to which refer the indices $I,J,K,L$), by the usual relations
\begin{equation}
	\label{PB_mec}
	\{A,B\}_{PB}=\frac{\partial A}{\partial\varepsilon_{IJ}}\frac{\partial B}{\partial\pi_{IJ}}-\frac{\partial A}{\partial\pi_{IJ}}\frac{\partial B}{\partial\varepsilon_{IJ}}.
\end{equation}

In our case, the dynamical variables are the real symmetric rank-two tensors, $\varepsilon_{IJ}$ and $\pi_{IJ}$ ($I,J=1,2,3$), which are canonically conjugate in the sense that their Poisson brackets are deemed to satisfy the following properties:
\begin{align}
	\label{rank2hamilton1}
	\{\varepsilon_{IJ},\pi_{KL}\}_{PB}         & =\delta_{IJKL},                \\
	\label{rank2hamilton2}
	\{\varepsilon_{IJ},\varepsilon_{KL}\}_{PB} & =\{\pi_{IJ},\pi_{KL}\}_{PB}= 0
\end{align}
where $\delta_{IJKL}$ is a $\delta$ ``tensor'', reflecting the real, symmetric, nature of the Poisson brackets~\cite{kijowskiCanonicalStructureClassical1976,*szczyrbaSymplecticStructureSet1976,*marsdenCovariantPoissonBrackets1986} and defined as a product of Kronecker $\delta$s. 
Very schematically, we may write 
\begin{equation}
	\delta_{IJKL}=\left\{
	\begin{array}{rcl}
		\delta_{IJ}\delta_{KL}\\
		\delta_{IJ}\delta_{LK}\\
		\delta_{JI}\delta_{LK}\\
		\delta_{JI}\delta_{KL}\\
	\end{array}
	\right.
\end{equation}
where each choice of the RHS corresponds to a choice of indices in the Poisson brackets.
This choice can be supplemented by any linear combination of these $\delta$s that enforces the symmetries of the tensors.  

We now wish to include as phase space coordinates, the components of the spin vector, $\bm{S}$. 
We follow reference~\cite{yangGeneralizationsClassicalPoisson1980} and the details are summarized in appendix~\ref{appendix1}. 
The ``generalized'' Poisson bracket, for the canonical variables of our system,  can be written as 
\begin{equation}
	\arraycolsep=1.4pt\def\arraystretch{2.2}
	\begin{array}{rcl}
		\displaystyle \{A,B\}_{PB}\equiv\frac{\partial A}{\partial\varepsilon_{IJ}}\frac{\partial B}{\partial\pi_{IJ}} & - & \displaystyle \frac{\partial A}{\partial\pi_{IJ}}\frac{\partial B}{\partial\varepsilon_{IJ}} \\
		& - & \displaystyle \frac{1}{\hbar}\epsilon_{IJK}S_I\frac{\partial A}{\partial S_J}\frac{\partial B}{\partial S_K},
	\end{array}
\end{equation}
where $\hbar$ is introduced to restore the physical dimensions of the Poisson bracket, since ${\bm S}$ is considered dimensionless, for the sake of simplicity.

It should be stressed that this $\hbar$ does not imply that any {\em quantum} effects are present, since the dynamics is fully classical. 
It is simply a bookkeeping device for a quantity that has the dimensions of angular momentum--i.e. of an area in phase space--and reflects the fact that the equation of motion for $S_I$ is of first order.
Quantum fluctuations will introduce the ``real'' $\hbar$. 

Using this Poisson bracket, we can obtain the equations of motion for the phase space variables:
\begin{equation}
	\arraycolsep=1.4pt\def\arraystretch{2.2}
	\begin{array}{rcl}
		\displaystyle \dot{\varepsilon}_{IJ} & = & \displaystyle \{\varepsilon_{IJ},{\cal H}\}_{PB}=\frac{\partial{\cal H}}{\partial\pi_{IJ}}                   \\
		\displaystyle \dot{\pi}_{IJ}         & = & \displaystyle \{\pi_{IJ},{\cal H}\}_{PB}=-\frac{\partial{\cal H}}{\partial\varepsilon_{IJ}}                  \\
		\displaystyle \dot{S}_I              & = & \{S_I,{\cal H}\}_{PB}=\displaystyle \frac{1}{\hbar}\varepsilon_{IJK}S_J\frac{\partial{\cal H}}{\partial S_K}
	\end{array}
\end{equation}
The consistency of this formalism can be checked by noting that these equations preserve the volume in phase space
\begin{equation}
	\frac{\partial\dot{\varepsilon}_{IJ}}{\partial\varepsilon_{IJ}}+\frac{\partial\dot{\pi}_{IJ}}{\partial\pi_{IJ}}+\frac{\partial\dot{S}_K}{\partial S_{K}}=0.
	\label{divergencephasespace}
\end{equation}
That the dynamics preserves the volume in phase space does not, of course, imply anything about whether the system thus defined is integrable or shows Hamiltonian chaos. 

The internal energy $U$, involving the mechanical energy for the deformed elastic medium~\cite{landauTheoryElasticity2008}, the magnetic energy, defined by the Zeeman term~\cite{skomskiSimpleModelsMagnetism2008} and the magneto-elastic energy~\cite{dutremoletdelacheisserieMagnetostrictionTheoryApplications1993}, that takes into account the interaction of the magnetic moment with the medium, takes the form
\begin{align}
	U=& \frac{V_0}{2}C_{IJKL}\varepsilon_{IJ}\varepsilon_{KL}-V_0\sigma^{\textrm{ext}}_{IJ}\varepsilon_{IJ}+B_{IJKL}^{(1)}\varepsilon_{IJ}S_KS_L\nonumber \\
	  & -\hbar\omega_IS_I,
	  \label{Energy_U}
\end{align}
where $C$ is the fully symmetric tensor, defining the elastic response, $\sigma^{\textrm{ext}}$ is the external stress tensor, ${\bm\omega}$ is the effective external magnetic field (expressed as a frequency) and $B^{(1)}$ is the fully symmetric linear magnetostriction tensor.

The ``kinetic'' term, containing the conjugate momenta, can be written, schematically, as 
\begin{equation}
	{\cal H}_\mathrm{kinetic}=\frac{1}{2}\pi_{IJ}M^{-1}_{IJKL}\pi_{KL}
	\label{Energy_T}
\end{equation}
where $M$ is a fully symmetric ``mass'' matrix, that describes the inertial response. 

For the case of isotropic materials, it is assuming that the $M$ tensor has the form given by Lamé, with only two characteristic constants:
\begin{equation}
	M_{IJKL}=M_0\delta_{IJ}\delta_{KL}+M_1\left(\delta_{IK}\delta_{JL}+\delta_{IL}\delta_{JK}\right).
\end{equation}
The tensors $C$ and $B$ are decomposed in the same way.

Consequently, the inverse of these tensors can then be deduced from
$M_{IJKL}M^{-1}_{IJMN}=\frac{1}{2}\left(\delta_{KM}\delta_{LN}+\delta_{KN}\delta_{LM}\right)$. Thus
\begin{equation}
	\arraycolsep=1.4pt\def\arraystretch{2.2}
	\begin{array}{rcl}
		M^{-1}_{IJKL} & = & \displaystyle\frac{-M_0}{2M_1(3M_0+2M_1)}\delta_{IJ}\delta_{KL}                       \\
		              & + & \displaystyle\frac{1}{4M_1}\left(\delta_{IK}\delta_{JL}+\delta_{IL}\delta_{JK}\right)
	\end{array}
\end{equation}
and the equations of motion become
\begin{equation}
	\label{eoms}
	\arraycolsep=1.4pt\def\arraystretch{2.2}
	\left\{
	\begin{array}{rcl}
		\dot{\epsilon}_{IJ} & = & M^{-1}_{IJKL}\pi_{KL}                                                             \\
		\dot{\pi}_{IJ}      & = & -V_0C_{IJKL}\epsilon_{KL}+V_0\sigma^{\textrm{ext}}_{IJ}-B_{IJKL}S_KS_L                  \\
		\dot{S}_I           & = & \varepsilon_{IJK}\left(\omega_J-\frac{2}{\hbar}B_{ABJC}^{(1)}\epsilon_{AB}S_C\right)S_K
	\end{array}
	\right.
\end{equation}
highlighting how the mechanical and magnetic subsystems are coupled. 

The last equation--as expected!--is a precession equation for the components of ${\bm S}$ around both the effective field ${\bm \omega}$ and an additional field, that depends on the strain tensor and the spin vector. 

In the following section we shall show how to solve these equations, in a way that preserves the symmetries of the extended phase space.

\section{Geometric integration\label{section3}}

Solving the coupled system of equations~\eqref{eoms} is the next step.

Since, in the previous section, we have shown that these equations describe a volume preserving transformation of the enlarged phase space, encompassing elastic {\em and} spin variables, it is natural to rewrite them, in terms of the action of a Liouville operator.
Therefore, we shall write eqs.\eqref{eoms} as
\begin{equation}
	\label{Liouville_system}
	\begin{array}{rcl}
		{\dot{\bm \varepsilon}} &=&{{\cal L}}_{\bm{\epsilon}}{\bm \varepsilon}, \\
		{\dot{\bm \pi}}         &=&{\cal L}_{\bm{\pi}}{\bm \pi},                \\
		{\dot{\bm S}}           &=&{\cal L}_{\bm S}{\bm S}.
	\end{array}
\end{equation}
where ${\cal L}$ is the Liouville operator. 
This formulation allows us to implement, manifestly, time-reversible, area preserving algorithms, for solving these equations numerically.

The general scheme is as follows: Consider an arbitrary function $f$ of the canonically conjugate variables of our many-body system. 
This function, $f({\bm\varepsilon},{\bm\pi},{\bm S})$, depends on the time $t$ implicitly; that is, through the dependence of $({\bm\varepsilon},{\bm\pi},{\bm S})$ on $t$.
The time derivative of $f$ is $\dot{f}$ such as
\begin{align}
	{\dot f}&=
	\dot{\varepsilon}_{IJ}\frac{\partial f}{\partial\varepsilon_{IJ}}+\dot{\pi}_{IJ}\frac{\partial f}{\partial\pi_{IJ}}+\dot{S}_{I}\frac{\partial f}{\partial S_{I}}\nonumber\\
	&\equiv {\cal L}f.
	\label{full_Liouville}
\end{align}
The last line defines the (total) Liouville operator 
\begin{equation}
	{\cal L}=\dot{\varepsilon}_{IJ}\frac{\partial }{\partial\varepsilon_{IJ}}+\dot{\pi}_{IJ}\frac{\partial }{\partial\pi_{IJ}}+\dot{S}_{I}\frac{\partial }{\partial S_{I}}.
\end{equation}
Equation \eqref{full_Liouville} can be integrated formally as an initial value problem to obtain $f$ at any time~:
\begin{equation}
	f({\bm\varepsilon}(t),{\bm\pi}(t),{\bm S}(t))=e^{{\cal L}t}f({\bm\varepsilon}(0),{\bm\pi}(0),{\bm S}(0)).
\end{equation}
It is not difficult to see that ${\cal L}={\cal L}_{\bm{\varepsilon}}+{\cal L}_{\bm{\pi}}+{\cal L}_{\bm{S}}$.
However, these single Liouville operators do not commute two-by-two, as the reader may easily check by computing ${\cal L}_{\bm{u}}{\cal L}_{\bm{v}}f-{\cal L}_{\bm{v}}{\cal L}_{\bm{u}}f\neq 0$ for any function $f$ and any combination $(u,v)$ of the individual Liouville operator ${\cal L}_{\bm{\varepsilon}},{\cal L}_{\bm{\pi}},{\cal L}_{\bm{S}}$.
This means that
\begin{equation}
	e^{{\cal L}t}=e^{{\cal L}_{\bm \varepsilon}t+{\cal L}_{\bm \pi}t+{\cal L}_{\bm S}t}\neq e^{{\cal L}_{\bm \varepsilon}t} e^{{\cal L}_{\bm \pi}t} e^{{\cal L}_{\bm S}t}.
\end{equation}
According to the Magnus expansion~\cite{blanesMagnusExpansionIts2009}, however,
it is always possible to express $e^{{\cal L}t}$ as a product of the individual operators at any given order in time, according to the so--called ``splitting method''~\cite{hairerGeometricNumericalIntegration2006}.
This ensures that the numerical algorithm preserves phase space volumes exactly. 

For instance, for a fixed timestep $\tau$, upon expanding up to the third order in time, the following sequence of products
\begin{align}
	\label{expansion1}
	e^{{\cal L}\tau}
	&=e^{{\cal L}_{\bm{S}}\frac{\tau}{4}}e^{{\cal L}_{\bm{\pi}}\frac{\tau}{2}}e^{{\cal L}_{\bm{S}}\frac{\tau}{4}}e^{{\cal L}_{\bm{\varepsilon}}\tau}e^{{\cal L}_{\bm{S}}\frac{\tau}{4}}e^{{\cal L}_{\bm{\pi}}\frac{\tau}{2}}e^{{\cal L}_{\bm{S}}\frac{\tau}{4}}+{\cal O}(\tau^3),
\end{align}
can be generated; this sequence is, in fact, one of six possible, that have the property of preserving the symplectic structure of the Poisson brackets. Therefore, any one of them can be chosen. 
The possible combinations are presented in table \ref{table:ST1}.
\begin{table}[!ht]
	\begin{center}
	\[
	\begin{array}{c|ccccccc}
		A&{\frac{\tau}{4}}&{\frac{\tau}{2}}&{\frac{\tau}{4}}&{\tau}&{\frac{\tau}{4}}&{\frac{\tau}{2}}&{\frac{\tau}{4}}
		\\\hline
		1&{\bm S}&{\bm\pi}&{\bm S}&{\bm\epsilon}&{\bm S}&{\bm\pi}&{\bm S}\\
		2&{\bm S}&{\bm\epsilon}&{\bm S}&{\bm\pi}&{\bm S}&{\bm\epsilon}&{\bm S}\\
		3&{\bm \pi}&{\bm\epsilon}&{\bm \pi}&{\bm S}&{\bm \pi}&{\bm\epsilon}&{\bm \pi}\\
		4&{\bm \pi}&{\bm S}&{\bm \pi}&{\bm\epsilon}&{\bm \pi}&{\bm S}&{\bm \pi}\\
		5&{\bm \epsilon}&{\bm\pi}&{\bm \epsilon}&{\bm S}&{\bm \epsilon}&{\bm\pi}&{\bm \epsilon}\\
		6&{\bm \epsilon}&{\bm S}&{\bm \epsilon}&{\bm\pi}&{\bm \epsilon}&{\bm S}&{\bm \epsilon}\\
		\end{array}
	\]
	\caption{Decomposition table of symplectic integrators\label{table:ST1}}	
	\end{center}
\end{table}
While these schemes are free from ``global'' errors, they are, of course, sensitive to ``local'' errors, due to the finite value of the timestep. It is, also, not at all obvious that all six can be implemented with comparable efficiency. It is, therefore, useful to study the numerical stability and efficiency of these different combinations, in particular, as former studies in molecular dynamics~\cite{batchoSpecialStabilityAdvantages2001} and magnetic molecular dynamics~\cite{beaujouanAnisotropicMagneticMolecular2012} showed, apparently, numerical differences between them.

A sampler of such a study is presented in appendix \ref{appendix2}.

It is important to keep in mind that the one--step evolution operators for ${\bm\varepsilon}$ and ${\bm\pi}$ describe shifts of the corresponding tensor components, whereas the one--step evolution operator for ${\bm S}$ describes  rotations.
In equations
\begin{align}
	e^{{\cal L}_{\varepsilon}\tau}({\bm\varepsilon}(t),{\bm\pi}(t),{\bm S}(t))&=({\bm\varepsilon}(t)+\tau{\dot{\bm\varepsilon}(t)},{\bm\pi}(t),{\bm S}(t))\\
	e^{{\cal L}_{\pi}\tau}({\bm\varepsilon}(t),{\bm\pi}(t),{\bm S}(0))&=({\bm\varepsilon}(t),{\bm\pi}(t)+\tau{\dot{\bm\pi}(t)},{\bm S}(t))\\
	e^{{\cal L}_{S}\tau}({\bm\varepsilon}(t),{\bm\pi}(t),{\bm S}(t))&=
	({\bm\varepsilon}(t),{\bm\pi}(t),\underbrace{{\bm S}(t+\tau)}_{R(\tau){\bm S}(t)})
\end{align}
where ${\bm S}(t+\tau)=R(\tau){\bm S}(t)$ is given by the Rodrigues' rotation formula~\cite{rodriguesLoisGeometriquesQui1840,*honerkampTheoreticalPhysicsClassical1993,*thibaudeauThermostattingAtomicSpin2012} for a spin vector around a given rotation vector ${\bm{\tilde{\omega}}}(t)$ where each of its components are $\tilde\omega_I(t)=\omega_I(t)-\frac{2}{\hbar}B^{(1)}_{JKLI}\varepsilon_{JK}(t)S_L(t)$.
These equations describe the phase space of one particle only. 
To describe the dynamics of a continuum, we must deduce the equations for many particles.

One way to generalize eqs.\eqref{eoms} for the case of many particles, labeled by an index, $i=1,\ldots,N$, according to the conventions of the previous sections, is the following
\begin{align}
	\label{eomsN1}
	\dot{\epsilon}^{i}_{IJ} & =  [M^{i}]^{-1}_{IJKL}\pi^{i}_{KL}\\
	\label{eomsN2}
	\dot{\pi}^{i}_{IJ}      & =  -V_0C^{i}_{IJKL}\epsilon^{i}_{KL}+V_0\sigma^{i\hspace{2pt}\textrm{ext}}_{IJ}-B^{i}_{IJKL}S^{i}_KS^{i}_L\\
	\label{eomsN3}
	\dot{S}^{i}_I           & =  \varepsilon_{IJK}\left(\omega^{i}_J-\frac{2}{\hbar}B^{(1)i}_{ABJC}\epsilon^{i}_{AB}S^{i}_C\right)S^{i}_K
\end{align}

The case of a staggered AF state is treated simply by letting $N=2$. 
In order to simplify the mechanical part further, we can impose additional conditions pertaining to the uniformity of the external stress, mechanical constants, mass matrices and magneto-elastic constants, at each site.
In what follows, we shall use the following {\em Ansatz}:
\begin{equation}
	\begin{array}{rcl}
		B^{(1)1}_{IJKL}&=&B^{(1)2}_{IJKL},\\
		C^{1}_{IJKL}&=&C^{2}_{IJKL},\\
		M^{1}_{IJKL}&=&M^{2}_{IJKL},\\
		\sigma^{1\hspace{2pt}\textrm{ext}}_{IJ}&=&\sigma^{2\hspace{2pt}\textrm{ext}}_{IJ}.
	\end{array}
\end{equation}
The conservative part of the precession contains a local field, which is modified to include the antiferromagnetic exchange between the sites and a single anisotropy axis ${\bm n}$ with an intensity $K$.
One has
\begin{align}
	\omega_I^i&=\frac{1}{\hbar}\sum_{<ij>}J^{ij}S_I^j+\frac{K}{\hbar} n_Js_J^in_I
\end{align}

Because of the exchange field, the Liouville operators for different spins do not commute either.
A global geometric integrator, implementing the approach of Omelyan~\cite{omelyanSymplecticAnalyticallyIntegrable2003}, which remains accurate up to third order in the timestep expansion must, therefore, be constructed.

For any given timestep, $\tau$, the corresponding expression for the evolution operator, reads
\begin{align}
	e^{{\cal L}_S\tau}&=e^{{\cal L}_{S^1}\frac{\tau}{2}}e^{{\cal L}_{S^2}\tau}e^{{\cal L}_{S^1}\frac{\tau}{2}}+{\cal O}(\tau^3).
\end{align}
For $N=2$, this operator is numerically identical to the operator, obtained by permuting the site indices, $1\leftrightarrow 2$. 

The same reasoning is applied for the Liouville operators for the elastic variables, that enter in equations \eqref{Liouville_system} and the corresponding global geometric integrators are constructed along the same lines.
The system is then integrated by following one of the schemes displayed in table \ref{table:ST1}.

With these tools, we can study a plethora of phenomena, that are sensitive to the  coupling of  magnetic, electric and elastic degrees of freedom.
In the following section we shall apply this formalism for studying ``switching'' effects of the magnetization in antiferromagnetic media.
The numerical accuracy ensured by the geometric integrators is necessary to describe picosecond switching times.

\section{Antiferromagnetic ultrafast switching under stress\label{section4}}

In this section, we shall apply the formalism constructed above, to describe how it is possible to generate and manipulate picosecond switching of the magnetization, induced by STTs, from a short pulse of electric current, in elastic media that exhibit staggered AF order.    
\begin{figure}[htp]
	\centering
	\includegraphics[width=\columnwidth]{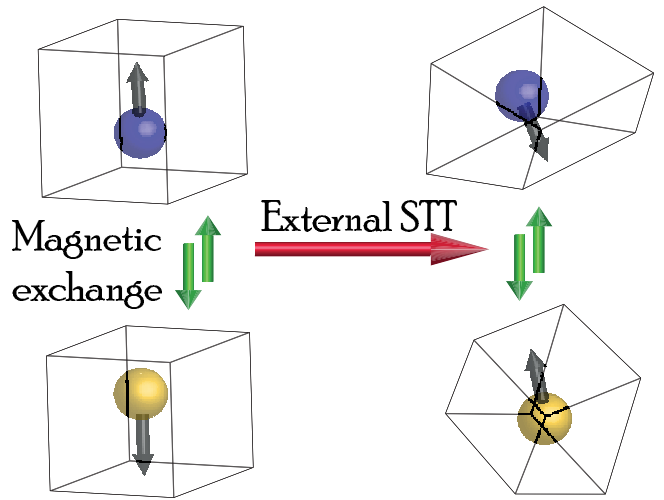}
	\caption{(Color online) Switching scheme for antiferromagnetic magneto-mechanically coupled toy model with external STT.}
	\label{DisplaySTT}
\end{figure}
Such a fast switching process in AFs is schematically displayed in figure \ref{DisplaySTT} and can be realized, in practice, by a femtosecond laser excitation of the magnetic moments that generate a field-induced STT on the two sub-lattices.
During the pulse, energy is transferred from conduction electrons to the sub-lattice magnetic moments via STT (this is how the electric current affects the magnetic response), and hence contributes to the exchange energy between the moments, since local moments can be canted noncollinearly.
The energy due to the strong exchange interaction between neighboring moments, which is commonly found in AFs, in particular due to magnetic anisotropy,  appears as an effective inertia to the motion, leading to the appearance of a time scale, much longer than that of the excitation pulse.
Afterwards, the system follows a natural path along the easy-plane to circumvent the unfavorable anisotropy barrier and finally relaxes to a new magnetic configuration.
The resulting dynamics is a switching of the two sub-lattices moments to the opposite direction through the easy-plane.
 
This process defines the ultrafast antiferromagnetic switching effect.
During the process, the Néel vector, that probes the difference between the magnetic moments of the two sub-lattices, acquires a net value that can be transferred as a so-called ``spin accumulation'', to an adjacent non-magnetic normal material in order to pump the produced spin current via scattering of electrons.
The reverse mechanism can be realized, as well.

What was not considered before is the possibility to enhance or inhibit such a switching, depending on tensile or compression effects, that are generated by an external stress, that couples to the internal strain, produced by the intrinsic magneto-elastic interaction in such materials, as depicted in figure \ref{DisplaySTT}.

In order to conform to the notation used in our previous studies~\cite{nussleCouplingMagnetoelasticLagrangians2019}, equation \eqref{eomsN3} is supplemented with a non-conservative part, labeled ${\bm T}$, on the RHS, which includes both a transverse damping and a damping-like STT torque
\begin{align}
	T_I^i&=\alpha\epsilon_{IJK}S_J^i\dot{S}_K^i+G\left(s_I^is_J^ip_J-p_Is_J^is_J^i\right).
\end{align}
The resulting equations of motion, are numerically integrated, using the approach of section \ref{section4}. 
It is noteworthy that, this equation, also, describes the ``backreaction'' of the magnetic response on the spin transfer torque. 
On the other hand, we do not consider how this backreaction affects the current pulse itself, that is assumed external. 

\begin{figure}[htp]
	\centering
	\includegraphics[width=\columnwidth]{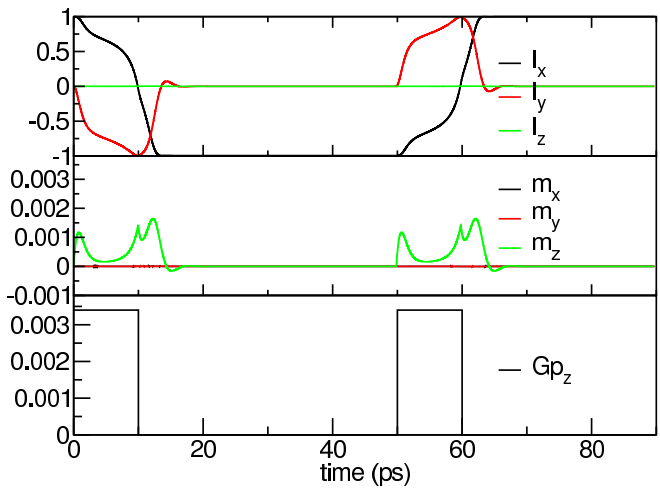}
	\caption{(Color online) Average magnetization (upper panel) and Néel order parameter components (middle panel) for uncoupled switching. \{$K=2\pi$rad.GHz, $\omega_E$=172.16 rad.THz, $M_s=5.10^5\mathrm{A.m}^{-1}$\}. Initial conditions: \{${\bm s}^1(0)=-{\bm s}^2(0)=\hat{\bm x}$\}. The lower panel displays the STT pulses. The figures agree with the reference\cite{chengUltrafastSwitchingAntiferromagnets2015} because the magnetoelastic constants are set to zero.}
	\label{Figure2}
\end{figure}

\begin{figure}[htp]
	\centering
	\includegraphics[width=\columnwidth]{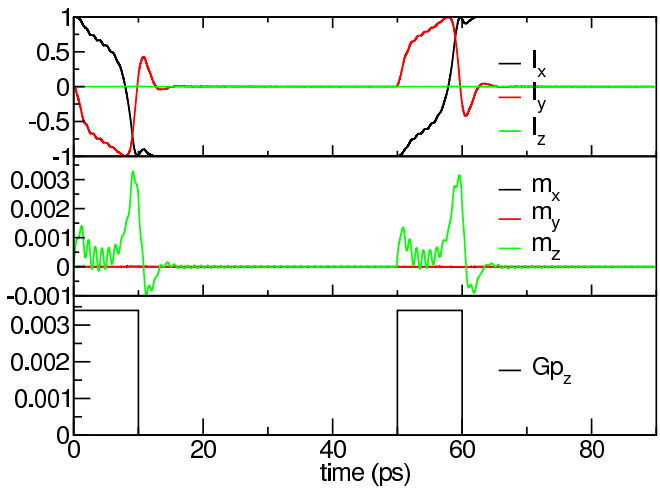}
	\caption{(Color online) Average magnetization (upper panel) and Néel order parameter components (middle panel) for a coupled switching. \{$K=2\pi$rad.GHz, $\omega_E$=172.16 rad.THz, $M_s=5.10^5\mathrm{A.m}^{-1}$\}. Initial conditions: \{${\bm s}^1(0)=-{\bm s}^2(0)=\hat{\bm x}$, $\epsilon^i_{IJ}(0)=0$\}. The lower panel displays the STT pulses. Here $B_0^{(1)}=7.7\mu_0 M_s^2$ and $B_1^{(1)}=-23\mu_0 M_s^2$.}
	\label{Figure3}
\end{figure}

In figures \ref{Figure2} and \ref{Figure3} we report the evolution of the average magnetization ${\bm m}\equiv \frac{1}{2}\left({\bm S}^1+{\bm S}^2\right)$ and the Néel vector ${\bm l}\equiv \frac{1}{2}\left({\bm S}^1-{\bm S}^2\right)$, in the presence as well as the absence of magnetoelastic coupling, for a moderate external stress of $\sigma_{xx}=-2\mu_0 M_s^2$.
We start the simulations using an initial configuration, where spins are aligned along the $\hat{\bm x}$-axis in an antiferromagnetic configuration and apply two $10$ ps electric pulses of $0.0034$ rad.THz intensity, each separated by $50$ ps, in the ${\bm{p}}={\hat{\bm{z}}}$-direction.
In addition to the exchange interaction, the spins are subjected to a global anisotropy, along the ${\bm{n}}={\hat{\bm{x}}}$-axis. 
The numerical value for $\alpha$ in the following studies is $0.005$.

In the absence of magnetoelastic coupling, our results are identical to those by Cheng {\em et al.}~\cite{chengUltrafastSwitchingAntiferromagnets2015} and to those we obtained under the same conditions in our previous work~\cite{nussleCouplingMagnetoelasticLagrangians2019}.
 
\begin{figure}[htp]
       \centering
       \includegraphics[width=\columnwidth]{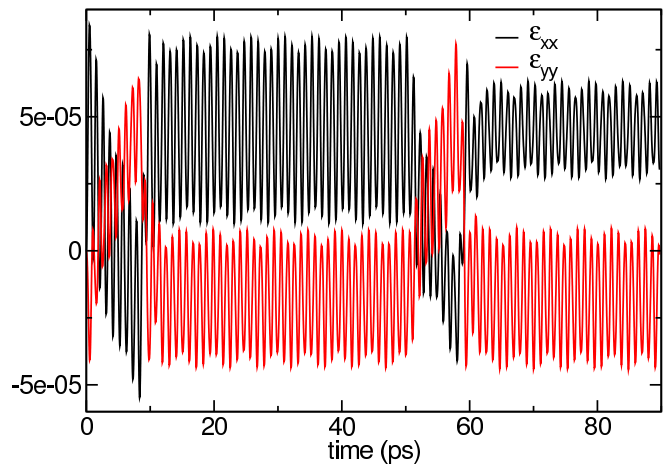}
       \caption{(Color online) Strain components as a function of time with the magnetoelastic constants turned on with parameters identical to figure \ref{Figure3}.}
       \label{Figure4}
\end{figure}

In the presence of magnetoelastic coupling, however, as the mechanical system is undamped, the results are quite different.
Because of the presence of a constant stress, a finite mass matrix and non-zero elastic constants, the mechanical system is expected to be oscillating freely, which is, indeed, observed in figure \ref{Figure4}, where we display the evolution of the strain components over time. 
Indeed, as one can see on figures \ref{Figure3} and \ref{Figure4} the characteristic oscillations of the mechanical system have repercussions on the Néel order parameter dynamics. 
This is even more striking for the spin accumulation which is more than doubled by the coupling to the strain ; here as well, the mechanical oscillations are reflected on the magnetic dynamics.
What is interesting is that these characteristic mechanical oscillations are expected to be related to the sound velocity of the medium~\cite{ogdenNonlinearElasticDeformations1997}, which is controlled by the constants $M_0$ and $M_1$.  In the present study, we have chosen $M_0=0$ and $M_1=10000$, but there should be a way to obtain these coefficients of the mass matrix more efficiently, in order to describe these oscillations more accurately from experimentally easily accessible data. 
This, however is not the focus of the present study and shall be studied more thoroughly elsewhere. 
As a consequence of such a stress induced strain, the switching time of the magnetization is slightly faster and the amplitude of the magnetization enhanced, depending on the values of the stress and magnetoelastic constants, as depicted in figure \ref{Figure3}. 
We observe that a compression in one direction becomes tensile in the other one and vice-versa depending on the sign of $B_1^{(1)}$. 
Conversely, if $B_1^{(1)}$ is of the opposite sign, a tensile stress would slow the switching process down. 

\begin{figure}[!htp]
	\centering
	\includegraphics[width=\columnwidth]{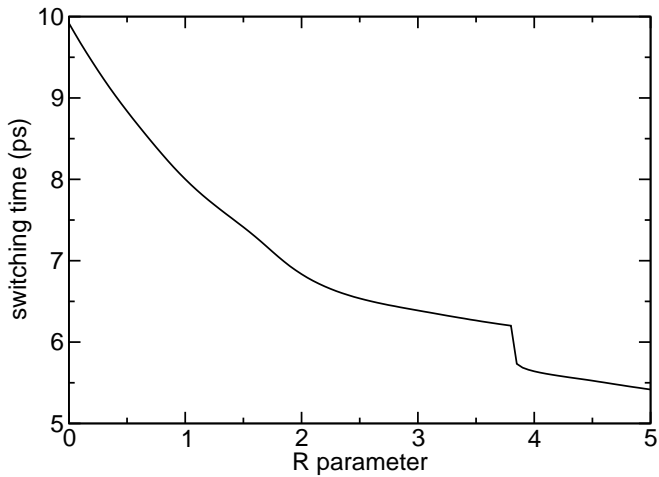}
	\caption{Switching time (in ps) as a function of varying $B_1^{(1)}$ over its natural value in NiO.}
	\label{Figure7}
\end{figure}

This leads to another interesting question which is how the switching time is affected by the values taken by either $B_0^{(1)}$ and $B_1^{(1)}$. 
We have verified that changing $B_0^{(1)}$ has no influence on it because this coefficient does not enter in the precession equation.
Its simply produces the Joule magnetostriction phenomena as anticipated before in reference \cite{dutremoletdelacheisserieMagnetostrictionTheoryApplications1993}.
However, changing $B_1^{(1)}$ has a strong influence.
The Figure \ref{Figure7} displays the switching time as a function of the values of varying $B_1^{(1)}$.
By defining $R$ as the ratio of $B_1^{(1)}$ divided by the its natural value in NiO, we perfom different simulations and report the time where the Neel vector $x$-component crosses over to negative values.
One can see that this time decreases by increasing $B_1^{(1)}$ which makes sense, as this is amplifying the effects of the tensile stress.
When $R$ becomes lower than $1$, the effect is opposite.
One aspect which is however surprising is that there is a quick drop of the switching time for approximately $4$ times the value of $B_1^{(1)}$ over its natural value in NiO.
This is due to the oscillations of the magnetization induced by the elastic coupling.
As this coupling becomes stronger, one of the mechanical oscillatory peaks dips below 0 and thus the switching time appears to drop in a discontinuous fashion. 
This is another interesting consequence to the interplay of the magnetization and the strain, that could be captured in real experiments.

\section{Discussion\label{discussion}}

In the present paper we have presented the details for a Hamiltonian framework that can describe the dynamics, as well as the coupling,  between  magnetism and elasticity by a careful definition of the corresponding dynamical variables in an extended phase-space. 

The magnetoelastic coupling is taken into account at the linear level only of the strain tensor, by the coefficients, $B_0$ and $B_1$, that can be derived either experimentally from magnetostriction constants~\cite{dutremoletdelacheisserieMagnetostrictionTheoryApplications1993} or by {\it ab initio} calculations~\cite{fahnleMagnetoelasticityFePossible2002}. 
It should be stressed that the elastic constants, $C_0$ and $C_1$ can be deduced from similar experiments, also. 

Notwithstanding the presence of only the linear term of the strain, it does describe non-trivial effects, that can be measured in real and numerical experiments, as we have shown. 
In particular, the ultrafast switching of the magnetization, induced by a damping-like STT in NiO, where both the duration and intensity can be modulated by appropriate internal strains, thanks to the magnetoelastic coupling.

It is, now, possible to set in perspective the limits of the present framework and how they may be overcome. 

The strain can be modulated, in return, by the induced magnetization process by the same magnetoelastic coupling, so as to give rise to  a modulated elastic wave, whose properties, in extended systems can leave a characteristic imprint, that would be interesting to look for. 

It should be kept in mind, though, that,  in real experiments, because amplitude fluctuations of the electric pulses are inevitable, small damping values are not favorable, since they  may easily lead to overshoot, which is amplified by the mechanical response. 
When Joule heating in the normal underlying metal is taken into account, a shorter pulse with stronger current intensity should be desirable, but a modulated undamped mechanical strain is then produced.
For the moment, a major conceptual issue is how to describe, in a consistent way, the  mechanical damping effects, that are induced by a differential thermal stress,  produced by such a Joule heating. 
We hope to return to this issue in future work. 
Hints may be gleaned from our previous work on the stochastic dynamics of the magnetization~\cite{tranchidaHierarchiesLandauLifshitzBlochEquations2018}. 

For the case of undamped dynamics we have developed  the mathematical and computational framework, that consistently takes into account elastic and magnetic degrees of freedom and that provides a starting point for designing future AFs devices for spintronics applications that combine both mechanical and magnetic responses and their interactions.
And we have, also, shown how to take into account  a particular class of damping terms, those that can describe spin--transfer--torque sources.

It is this framework that is the starting point for taking into account the stochastic effects mentioned above. 

\section*{Acknowledgments\label{acknowledgments}}

All authors contributed equally to the manuscript. T.N. acknowledges financial support through a joint doctoral fellowship ``CEA--Région Centre'' under grant number 201600110872.

\appendix
\section{A model for precession through anticommuting variables\label{appendix1}}

We review  some issues when dealing with classical spin variables and explain some limitations which are met by using the framework of commuting variables~\cite{brinkLocallySupersymmetricReparametrization1976,*brinkLagrangianFormulationClassical1977}.
A common starting point to build precessional models is choosing the appropriate representation of the ${\mathfrak{su}}(2)$ algebra.

In the  Heisenberg picture the Ehrenfest theorem implies that  the equations of motion for the average of the spin operator $\hat{\bm S}$ components read
\begin{equation}
	\imath\hbar\frac{d\langle \hat{\bm S}\rangle}{dt}=\langle[\hat{\bm S},{\cal H}] \rangle
\end{equation}
where ${\cal H}$ is the Zeeman Hamiltonian operator for an external magnetic field $\bm{H}$ such that
\begin{equation}
	{\cal H}=-g \mu_0\mu_B\hat{\bm S}\cdot\bm{H}.
\end{equation}
From the commutation relations $[\hat{S}_i,\hat{S}_j]=\imath\epsilon_{ijk}\hat{S}_k$, one can quite straightforwardly show that
\begin{equation}
	\frac{d\langle\hat{\bm S}\rangle}{dt}=\frac{g\mu_B\mu_0}{\hbar}\langle\hat{\bm S}\rangle\times\bm{H},
\end{equation}
which corresponds to the well-known Larmor precession of the expectation value  of the spin in an external magnetic field $\bm{H}$.

Interestingly, a different--but equivalent--approach can be followed, by considering the Majorana representation, in terms of anticommuting variables, $\xi_I$, 
$\xi_I\xi_J+\xi_J\xi_I=0$
~\cite{aliceaNewDirectionsPursuit2012} of the vector $\bm{S}$:
\begin{equation}
	S_I=-\frac{\imath}{2}\epsilon_{IJK}\xi_J\xi_K,
	\label{Majorana_fermions}
\end{equation}
which can easily be shown to commute, as $S_IS_J=0$. This last relation is, of course, not satisfactory. 
It can be easily avoided, however, by imposing that the $\xi_I$ satisfy the anticommutation relations $\{\xi_I,\xi_J\}=\delta^{ij}$--that they generate a Clifford algebra (up to a constant normalization)~\cite{lounestoCliffordAlgebrasSpinors2003}. 

These relations imply that the $S_I$ define a spin--$\frac{1}{2}$ representation of the $\mathfrak{su}(2)$ algebra. Higher spin representations can be defined by multilinear combinations~\cite{berezinParticleSpinDynamics1977}, that are relevant for describing magnetic properties of composite objects; since we can work with the $S_I$ instead of the $\xi_I$, however, this complication will not affect us here.

This representation highlights that the spin degrees of freedom, $S_I$, are, in fact, ``composite" objects and that the ``fundamental degrees of freedom'' are the $\xi_I$. 
Therefore, it is useful to develop the description of the dynamics, directly, in terms of the $\xi_I$ themselves. 
We shall recall the salient features below.

The Poisson brackets for the anticommuting variables $\xi_I$ are related to those of $S_I$, in order that the dynamics be, indeed, equivalent, in a way that was set forth many years ago, through the construction of a corresponding  graded Poisson bracket, which generalizes Poisson brackets from manifolds to super-manifolds~\cite{casalbuoniQuantizationSystemsAnticommuting1976}.

In terms of any functions of anti-commuting variables $\bm{\xi}$ this bracket reads
\begin{equation}
	\label{PBFxi}
	\left\{f({\bm\xi}),g({\bm\xi})\right\}_{\textrm{PB}}\equiv\frac{\imath}{\hbar} f({\bm\xi})\frac{\overleftarrow{\partial}}{\partial\xi_K}\frac{\overrightarrow{\partial}}{\partial\xi_K}g({\bm\xi}),
\end{equation}
with the corresponding definition of the left and right derivative of any function of the anti-commuting variables.
One can show that this bracket, also known as the ``antibracket'' of any two functions on a flat supermanifold, satisfies all necessary properties for a graded bracket, namely (graded) Leibniz rule, (graded) anti-symmetry and (graded) Jacobi identity~\cite{gomisAntibracketAntifieldsGaugetheory1995,*kanatchikovFieldTheoreticGeneralizations1997}. 

By taking this graded Poisson bracket for any two of these anti-commuting variables, we get
\begin{equation}
	\{\xi_I,\xi_J\}_{PB}=\frac{\imath}{\hbar}\delta_{IJ}.
\end{equation}
This implies that any two such variables are  canonically conjugate, since  $\{\xi_I,-\imath\hbar\xi_J\}_{PB}=\delta_{IJ}$ and $\pi_I\equiv -\imath\hbar\xi_I$ defines the canonical conjugate~\cite{casalbuoniClassicalMechanicsBosefermi1976,berezinParticleSpinDynamics1977}.

By using the Grassmann properties of ${\bm\xi}$, one proves that
\begin{equation}
	\{S_I,S_J\}_{PB}=\frac{1}{\hbar}\epsilon_{IJK}S_K
\end{equation}
which is a consistency check that the $S_I$ as defined in \eqref{Majorana_fermions} do realize a representation of the rotation group\cite{varshalovichQuantumTheoryAngular1988}. 

In eq.\eqref{PBFxi}, if the functions of the $\bm{\xi}$ are chosen to contain only quadratic terms, then one can identify the previous bracket as a regular Poisson bracket on a Riemannian manifold for the commuting variables $\bm{S}$
\begin{equation}
	\arraycolsep=1.4pt\def\arraystretch{2.2}
	\begin{array}{rcl}
		\displaystyle \{f({\bm S}),g({\bm S})\}_{PB}=\frac{\imath}{\hbar}\frac{\partial f}{\partial S_I}S_I\frac{\overleftarrow{\partial}}{\partial\xi_K}\frac{\overrightarrow{\partial}}{\partial\xi_K}S_J\frac{\partial g}{\partial S_J} \\
		\displaystyle =\{S_I,S_J\}_{PB}\frac{\partial f}{\partial S_I}\frac{\partial g}{\partial S_J} =\frac{1}{\hbar}\epsilon_{IJK}S_K\frac{\partial f}{\partial S_I}\frac{\partial g}{\partial S_J},
	\end{array}
\end{equation}
which is, precisely, the ``spinning part'' of the bracket introduced by Yang and Hirschfelder~\cite{yangGeneralizationsClassicalPoisson1980} for magnetized fluid dynamics, which reads as
\begin{equation}
	\{A,B\}_{PB}\equiv\frac{\partial A}{\partial q_{I}}\frac{\partial B}{\partial p_{I}}-\frac{\partial A}{\partial p_{I}}\frac{\partial B}{\partial q_{I}}-\frac{1}{\hbar}\epsilon_{IJK}S_I\frac{\partial A}{\partial S_J}\frac{\partial B}{\partial S_K}.
\end{equation}
Similar expressions have been found already by Casalbuoni a long time ago~\cite{casalbuoniClassicalMechanicsBosefermi1976} without having attracted  the attention they deserve.
Following the inverse path of canonical quantization, we use this graded Poisson Bracket on any commuting quantity, which allows us to compute directly
\begin{equation}
	\{\bm{S},{\cal H}\}_{PB}=\frac{g\mu_B\mu_0}{\hbar}\bm{S}\times\bm{H}
\end{equation}
thereby highlighting that the description of $\bm{S}$ in terms of $\bm{\xi}$ is an equivalent description of the dynamics. 

For any time-dependent commuting functions of ${\bm S}(t)$, we can use this Poisson bracket to deduce a Liouville equation
\begin{equation}
	\frac{dF({\bm S}(t))}{dt}=\{F({\bm S}(t)),{\cal H}\}_{PB}.
\end{equation}
Consequently, the equations of motion for the variables ${\bm \xi}$ read much more simple expressions as
\begin{equation}
	\label{eomxi}
	\frac{d{\bm \xi}(t)}{dt}=\{{\bm\xi},{\cal H}\}_{PB}=\frac{\imath}{\hbar}\frac{\overrightarrow{\partial}{\cal H}}{\partial{\bm\xi}},
\end{equation}
which are known to form a non-relativistic pseudoclassical mechanics\cite{casalbuoniClassicalMechanicsBosefermi1976}.

One reason why using the representation of $\bm{S}$ in terms of $\bm{\xi}$ is useful is that it is intrinsically difficult to build a Lagrangian model for the commuting spin variable ${\bm S}$, since its canonically conjugate variable cannot be unambiguously identified. 
As is, by now, well known, the conjugate of the dynamical variable ${\bm \xi}$ is proportional to itself~\cite{casalbuoniRelativelySupersymmetries1976}.
Therefore, the dynamics of spinning degrees of freedom can be described, either through a vector of commuting variables on a curved manifold, or by a vector of anti-commuting variables on a flat--though non--Riemannian--manifold.
Finally, it has been found that the Majorana-fermion representation of $\frac{1}{2}$-spin operators is also a powerful tool to straightforwardly compute spin-spin correlators~\cite{shnirmanSpinSpinCorrelatorsMajorana2003,*maoSpinDynamicsMajorana2003,*schadMajoranaRepresentationDissipative2015}, which represents an advantage for computing magnetic response functions in many-body systems.   

\section{Numerical accuracy of the splitting algorithms\label{appendix2}}

The accuracy of the numerical schemes represented in table \ref{table:ST1} depends on the relative amplitude of the velocity terms $(\dot{\varepsilon},\dot{\pi},\dot{S})$, and can be monitored by checking the stability of equation \eqref{divergencephasespace} over time~\cite{hairerGeometricNumericalIntegration2006,blanesMagnusExpansionIts2009}.

\begin{figure}[htbp]
	\centering
	\includegraphics[width=\columnwidth]{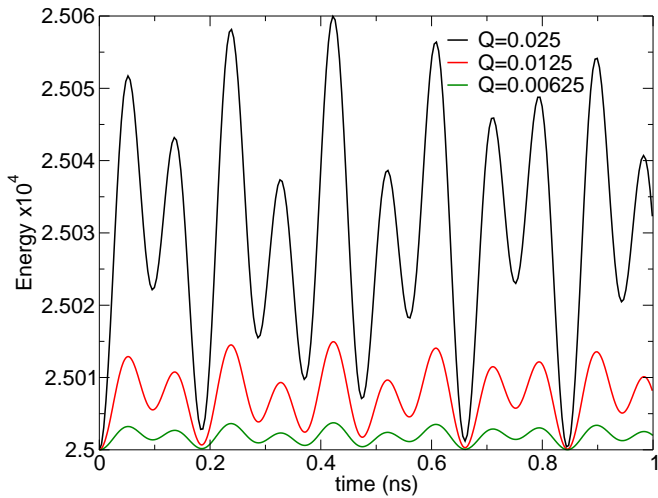}
	\caption{(Color online) Single-site total energy as a function of both time and variable timestep $Q$. Conditions of the simulation are expressed in reduced units: $2C_0/\mu_0 M_s^2=5.1\times 10^5$, $2C_1/\mu_0 M_s^2=3.5\times 10^5$, $M_1V_0\mu_0 M_s^2/2\hbar^2=1000$, ${\bm\omega}_{DC}=(0,0,2\pi)$ rad.GHz, $\pi_{11}(0)=1$, ${\bm{s}}(0)=(1,0,0)$. All the other parameters not reported, included initial conditions are zero.}
	\label{FigureA2_1}
\end{figure}
Figure \ref{FigureA2_1} displays the total energy in one domain (given by the sum of eqs.\eqref{Energy_U} and \eqref{Energy_T}) upon varying the numerical precision.
The splitting algorithm considered corresponds to the label $A= 1$ in table \ref{table:ST1}.
The numerical precision is controlled  by the ``quality factor $Q$'', chosen at the beginning of the simulation, which produces variable timesteps $\tau$ according to the relation $\tau=Q/\|\bf{\omega}\|$.
Simulation conditions produce a total energy equal to $\pi_{11}^2(0)/4M_1$, which has to stay constant over time.
We observe that this is, indeed,  the case, whatever  the value of the  quality factor $Q$. We remark that, as $Q\to 0$, the
variations about the average value of the energy are suppressed, as should be expected. 

\begin{figure}[!htbp]
	\centering
	\includegraphics[width=\columnwidth]{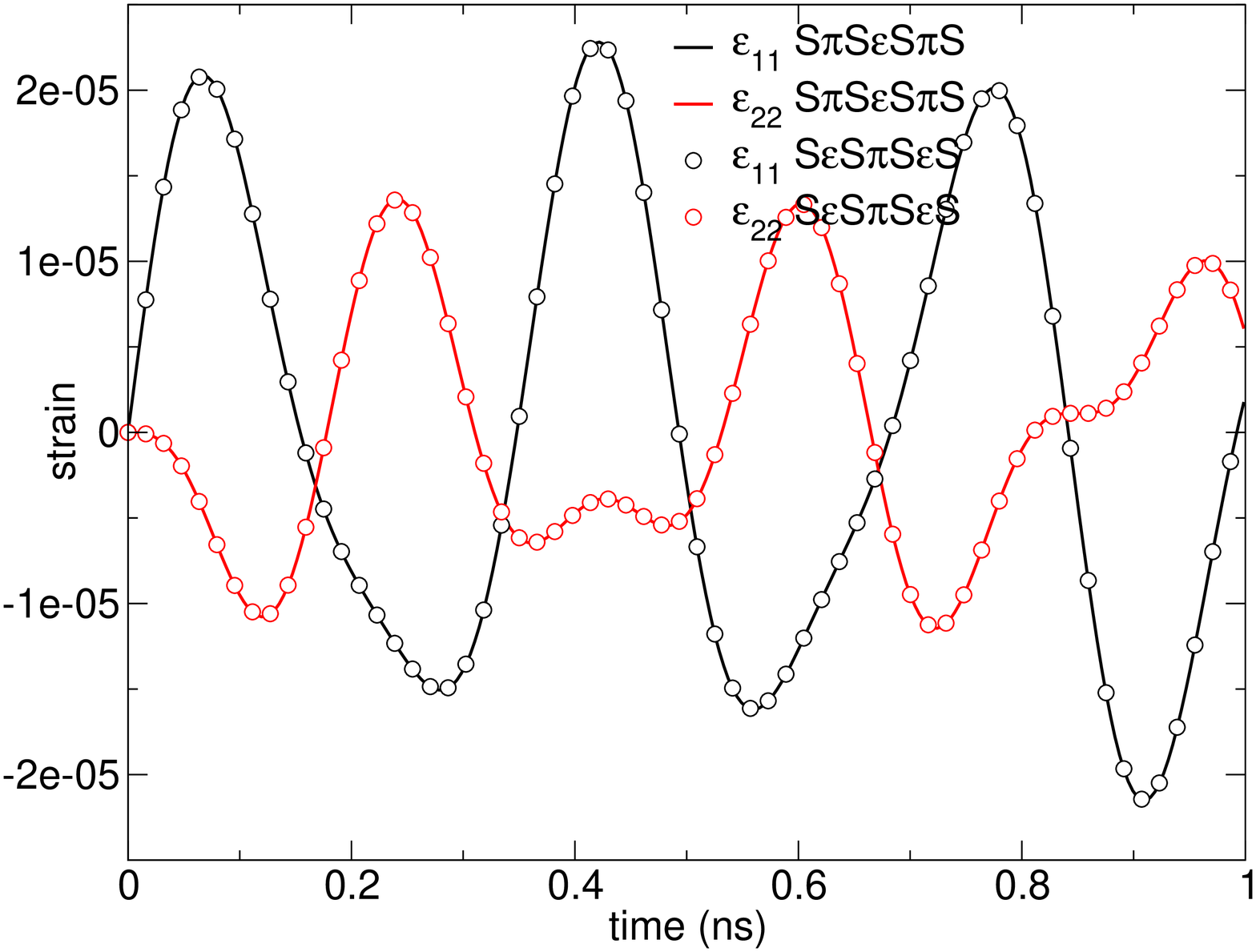}
	\caption{(Color online) Single-site strain components as a function of time for various numerical schemes. Conditions of the simulation are identical to those for figure \ref{FigureA2_1} and the results are produced for $Q=0.0025$ only.}
	\label{FigureA2_2}
\end{figure}

Figure \ref{FigureA2_2} displays some of the non-vanishing components of the strain tensor over time (here $\varepsilon_{22}(t)=\varepsilon_{33}(t)$ and  this third component is not reported), when using different splitting algorithms, among those displayed in Table \ref{table:ST1}.
We observe less than $1\%$ of numerical relative difference between the two algorithms on the strain and its conjugate variables, and no difference (up to the machine precision) on the magnetization with a ``coarse'' quality factor $Q$, which cannot be detected by visual inspection on the figure.
This difference falls to $0.1\%$ when the quality factor is divided by $4$.
The same procedure can be repeated for all the splitting combinations in Table \ref{table:ST1}, the conclusions previously drawn apply, also, for the magnetization, strain and its conjugate variable, depending on the frequency of appearance of the splitted operator.

As expected, once a splitting algorithm is selected, the more frequently an operator is evaluated, the smaller the error, though, of course, the time required, also, grows. 
This opens the possibility to select optimally a proper splitting algorithm, depending on the relative intensity of $\dot{\varepsilon}_{ij},\dot{\pi}_{ij},{\dot{S}}_i$ over time.     

\begin{thebibliography}{69}%
	\makeatletter
	\providecommand \@ifxundefined [1]{%
	 \@ifx{#1\undefined}
	}%
	\providecommand \@ifnum [1]{%
	 \ifnum #1\expandafter \@firstoftwo
	 \else \expandafter \@secondoftwo
	 \fi
	}%
	\providecommand \@ifx [1]{%
	 \ifx #1\expandafter \@firstoftwo
	 \else \expandafter \@secondoftwo
	 \fi
	}%
	\providecommand \natexlab [1]{#1}%
	\providecommand \enquote  [1]{``#1''}%
	\providecommand \bibnamefont  [1]{#1}%
	\providecommand \bibfnamefont [1]{#1}%
	\providecommand \citenamefont [1]{#1}%
	\providecommand \href@noop [0]{\@secondoftwo}%
	\providecommand \href [0]{\begingroup \@sanitize@url \@href}%
	\providecommand \@href[1]{\@@startlink{#1}\@@href}%
	\providecommand \@@href[1]{\endgroup#1\@@endlink}%
	\providecommand \@sanitize@url [0]{\catcode `\\12\catcode `\$12\catcode
	  `\&12\catcode `\#12\catcode `\^12\catcode `\_12\catcode `\%12\relax}%
	\providecommand \@@startlink[1]{}%
	\providecommand \@@endlink[0]{}%
	\providecommand \url  [0]{\begingroup\@sanitize@url \@url }%
	\providecommand \@url [1]{\endgroup\@href {#1}{\urlprefix }}%
	\providecommand \urlprefix  [0]{URL }%
	\providecommand \Eprint [0]{\href }%
	\providecommand \doibase [0]{http://dx.doi.org/}%
	\providecommand \selectlanguage [0]{\@gobble}%
	\providecommand \bibinfo  [0]{\@secondoftwo}%
	\providecommand \bibfield  [0]{\@secondoftwo}%
	\providecommand \translation [1]{[#1]}%
	\providecommand \BibitemOpen [0]{}%
	\providecommand \bibitemStop [0]{}%
	\providecommand \bibitemNoStop [0]{.\EOS\space}%
	\providecommand \EOS [0]{\spacefactor3000\relax}%
	\providecommand \BibitemShut  [1]{\csname bibitem#1\endcsname}%
	\let\auto@bib@innerbib\@empty
	\bibitem [{\citenamefont {Radu}\ \emph {et~al.}(2011)\citenamefont {Radu},
	  \citenamefont {Vahaplar}, \citenamefont {Stamm}, \citenamefont {Kachel},
	  \citenamefont {Pontius}, \citenamefont {D{\"u}rr}, \citenamefont {Ostler},
	  \citenamefont {Barker}, \citenamefont {Evans}, \citenamefont {Chantrell}
	  \emph {et~al.}}]{raduTransientFerromagneticlikeState2011}%
	  \BibitemOpen
	  \bibfield  {author} {\bibinfo {author} {\bibfnamefont {I.}~\bibnamefont
	  {Radu}}, \bibinfo {author} {\bibfnamefont {K.}~\bibnamefont {Vahaplar}},
	  \bibinfo {author} {\bibfnamefont {C.}~\bibnamefont {Stamm}}, \bibinfo
	  {author} {\bibfnamefont {T.}~\bibnamefont {Kachel}}, \bibinfo {author}
	  {\bibfnamefont {N.}~\bibnamefont {Pontius}}, \bibinfo {author} {\bibfnamefont
	  {H.}~\bibnamefont {D{\"u}rr}}, \bibinfo {author} {\bibfnamefont
	  {T.}~\bibnamefont {Ostler}}, \bibinfo {author} {\bibfnamefont
	  {J.}~\bibnamefont {Barker}}, \bibinfo {author} {\bibfnamefont
	  {R.}~\bibnamefont {Evans}}, \bibinfo {author} {\bibfnamefont
	  {R.}~\bibnamefont {Chantrell}},  \emph {et~al.},\ }\href {\doibase
	  10.1038/nature09901} {\bibfield  {journal} {\bibinfo  {journal} {Nature}\
	  }\textbf {\bibinfo {volume} {472}},\ \bibinfo {pages} {205} (\bibinfo {year}
	  {2011})}\BibitemShut {NoStop}%
	\bibitem [{\citenamefont {Jungwirth}\ \emph {et~al.}(2016)\citenamefont
	  {Jungwirth}, \citenamefont {Marti}, \citenamefont {Wadley},\ and\
	  \citenamefont {Wunderlich}}]{jungwirthAntiferromagneticSpintronics2016}%
	  \BibitemOpen
	  \bibfield  {author} {\bibinfo {author} {\bibfnamefont {T.}~\bibnamefont
	  {Jungwirth}}, \bibinfo {author} {\bibfnamefont {X.}~\bibnamefont {Marti}},
	  \bibinfo {author} {\bibfnamefont {P.}~\bibnamefont {Wadley}}, \ and\ \bibinfo
	  {author} {\bibfnamefont {J.}~\bibnamefont {Wunderlich}},\ }\href {\doibase
	  10.1038/nnano.2016.18} {\bibfield  {journal} {\bibinfo  {journal} {Nature
	  Nanotechnology}\ }\textbf {\bibinfo {volume} {11}},\ \bibinfo {pages} {231}
	  (\bibinfo {year} {2016})}\BibitemShut {NoStop}%
	\bibitem [{\citenamefont {{\v Z}elezn{\'y}}\ \emph {et~al.}(2018)\citenamefont
	  {{\v Z}elezn{\'y}}, \citenamefont {Wadley}, \citenamefont {Olejn{\'i}k},
	  \citenamefont {Hoffmann},\ and\ \citenamefont
	  {Ohno}}]{zeleznySpinTransportSpin2018}%
	  \BibitemOpen
	  \bibfield  {author} {\bibinfo {author} {\bibfnamefont {J.}~\bibnamefont {{\v
	  Z}elezn{\'y}}}, \bibinfo {author} {\bibfnamefont {P.}~\bibnamefont {Wadley}},
	  \bibinfo {author} {\bibfnamefont {K.}~\bibnamefont {Olejn{\'i}k}}, \bibinfo
	  {author} {\bibfnamefont {A.}~\bibnamefont {Hoffmann}}, \ and\ \bibinfo
	  {author} {\bibfnamefont {H.}~\bibnamefont {Ohno}},\ }\href {\doibase
	  10.1038/s41567-018-0062-7} {\bibfield  {journal} {\bibinfo  {journal} {Nature
	  Physics}\ }\textbf {\bibinfo {volume} {14}},\ \bibinfo {pages} {220}
	  (\bibinfo {year} {2018})}\BibitemShut {NoStop}%
	\bibitem [{\citenamefont {Ahmad}\ \emph {et~al.}(2015)\citenamefont {Ahmad},
	  \citenamefont {Atulasimha},\ and\ \citenamefont
	  {Bandyopadhyay}}]{ahmadReversibleStraininducedMagnetization2015}%
	  \BibitemOpen
	  \bibfield  {author} {\bibinfo {author} {\bibfnamefont {H.}~\bibnamefont
	  {Ahmad}}, \bibinfo {author} {\bibfnamefont {J.}~\bibnamefont {Atulasimha}}, \
	  and\ \bibinfo {author} {\bibfnamefont {S.}~\bibnamefont {Bandyopadhyay}},\
	  }\href {\doibase 10.1038/srep18264} {\bibfield  {journal} {\bibinfo
	  {journal} {Scientific Reports}\ }\textbf {\bibinfo {volume} {5}},\ \bibinfo
	  {pages} {18264} (\bibinfo {year} {2015})}\BibitemShut {NoStop}%
	\bibitem [{\citenamefont {Yan}\ \emph {et~al.}(2019)\citenamefont {Yan},
	  \citenamefont {Feng}, \citenamefont {Shang}, \citenamefont {Wang},
	  \citenamefont {Hu}, \citenamefont {Wang}, \citenamefont {Zhu}, \citenamefont
	  {Wang}, \citenamefont {Chen}, \citenamefont {Hua}, \citenamefont {Lu},
	  \citenamefont {Wang}, \citenamefont {Qin}, \citenamefont {Guo}, \citenamefont
	  {Zhou}, \citenamefont {Leng}, \citenamefont {Liu}, \citenamefont {Jiang},
	  \citenamefont {Coey},\ and\ \citenamefont
	  {Liu}}]{yanPiezoelectricStraincontrolledAntiferromagnetic2019}%
	  \BibitemOpen
	  \bibfield  {author} {\bibinfo {author} {\bibfnamefont {H.}~\bibnamefont
	  {Yan}}, \bibinfo {author} {\bibfnamefont {Z.}~\bibnamefont {Feng}}, \bibinfo
	  {author} {\bibfnamefont {S.}~\bibnamefont {Shang}}, \bibinfo {author}
	  {\bibfnamefont {X.}~\bibnamefont {Wang}}, \bibinfo {author} {\bibfnamefont
	  {Z.}~\bibnamefont {Hu}}, \bibinfo {author} {\bibfnamefont {J.}~\bibnamefont
	  {Wang}}, \bibinfo {author} {\bibfnamefont {Z.}~\bibnamefont {Zhu}}, \bibinfo
	  {author} {\bibfnamefont {H.}~\bibnamefont {Wang}}, \bibinfo {author}
	  {\bibfnamefont {Z.}~\bibnamefont {Chen}}, \bibinfo {author} {\bibfnamefont
	  {H.}~\bibnamefont {Hua}}, \bibinfo {author} {\bibfnamefont {W.}~\bibnamefont
	  {Lu}}, \bibinfo {author} {\bibfnamefont {J.}~\bibnamefont {Wang}}, \bibinfo
	  {author} {\bibfnamefont {P.}~\bibnamefont {Qin}}, \bibinfo {author}
	  {\bibfnamefont {H.}~\bibnamefont {Guo}}, \bibinfo {author} {\bibfnamefont
	  {X.}~\bibnamefont {Zhou}}, \bibinfo {author} {\bibfnamefont {Z.}~\bibnamefont
	  {Leng}}, \bibinfo {author} {\bibfnamefont {Z.}~\bibnamefont {Liu}}, \bibinfo
	  {author} {\bibfnamefont {C.}~\bibnamefont {Jiang}}, \bibinfo {author}
	  {\bibfnamefont {M.}~\bibnamefont {Coey}}, \ and\ \bibinfo {author}
	  {\bibfnamefont {Z.}~\bibnamefont {Liu}},\ }\href {\doibase
	  10.1038/s41565-018-0339-0} {\bibfield  {journal} {\bibinfo  {journal} {Nature
	  Nanotechnology}\ }\textbf {\bibinfo {volume} {14}},\ \bibinfo {pages} {131}
	  (\bibinfo {year} {2019})}\BibitemShut {NoStop}%
	\bibitem [{\citenamefont
	  {Fiebig}(2005)}]{fiebigRevivalMagnetoelectricEffect2005}%
	  \BibitemOpen
	  \bibfield  {author} {\bibinfo {author} {\bibfnamefont {M.}~\bibnamefont
	  {Fiebig}},\ }\href {\doibase 10.1088/0022-3727/38/8/R01} {\bibfield
	  {journal} {\bibinfo  {journal} {Journal of Physics D: Applied Physics}\
	  }\textbf {\bibinfo {volume} {38}},\ \bibinfo {pages} {R123} (\bibinfo {year}
	  {2005})}\BibitemShut {NoStop}%
	\bibitem [{\citenamefont {Atulasimha}\ and\ \citenamefont
	  {Bandyopadhyay}(2010)}]{atulasimhaBennettClockingNanomagnetic2010}%
	  \BibitemOpen
	  \bibfield  {author} {\bibinfo {author} {\bibfnamefont {J.}~\bibnamefont
	  {Atulasimha}}\ and\ \bibinfo {author} {\bibfnamefont {S.}~\bibnamefont
	  {Bandyopadhyay}},\ }\href {\doibase 10.1063/1.3506690} {\bibfield  {journal}
	  {\bibinfo  {journal} {Applied Physics Letters}\ }\textbf {\bibinfo {volume}
	  {97}},\ \bibinfo {pages} {173105} (\bibinfo {year} {2010})}\BibitemShut
	  {NoStop}%
	\bibitem [{\citenamefont {{Du Tr{\'e}molet de
	  Lacheisserie}}(1993)}]{dutremoletdelacheisserieMagnetostrictionTheoryApplications1993}%
	  \BibitemOpen
	  \bibfield  {author} {\bibinfo {author} {\bibfnamefont {E.}~\bibnamefont {{Du
	  Tr{\'e}molet de Lacheisserie}}},\ }\href@noop {} {{\selectlanguage
	  {English}\emph {\bibinfo {title} {Magnetostriction: Theory and Applications
	  of Magnetoelasticity}}}}\ (\bibinfo  {publisher} {{CRC Press}},\ \bibinfo
	  {address} {{Boca Raton}},\ \bibinfo {year} {1993})\BibitemShut {NoStop}%
	\bibitem [{\citenamefont
	  {Maugin}(1988)}]{mauginContinuumMechanicsElectromagnetic1988}%
	  \BibitemOpen
	  \bibfield  {author} {\bibinfo {author} {\bibfnamefont {G.~A.}\ \bibnamefont
	  {Maugin}},\ }\href@noop {} {{\selectlanguage {English}\emph {\bibinfo {title}
	  {Continuum Mechanics of Electromagnetic Solids}}}},\ \bibinfo {series} {North
	  {{Holland}} Series in Applied Mathematics and Mechanics}\ No.~\bibinfo
	  {number} {33}\ (\bibinfo  {publisher} {{North Holland}},\ \bibinfo {address}
	  {{Amsterdam}},\ \bibinfo {year} {1988})\BibitemShut {NoStop}%
	\bibitem [{\citenamefont
	  {Kittel}(1949)}]{kittelPhysicalTheoryFerromagnetic1949}%
	  \BibitemOpen
	  \bibfield  {author} {\bibinfo {author} {\bibfnamefont {C.}~\bibnamefont
	  {Kittel}},\ }\href {\doibase 10.1103/RevModPhys.21.541} {\bibfield  {journal}
	  {\bibinfo  {journal} {Reviews of Modern Physics}\ }\textbf {\bibinfo {volume}
	  {21}},\ \bibinfo {pages} {541} (\bibinfo {year} {1949})}\BibitemShut
	  {NoStop}%
	\bibitem [{\citenamefont {Buiron}\ \emph {et~al.}(1999)\citenamefont {Buiron},
	  \citenamefont {Hirsinger},\ and\ \citenamefont
	  {Billardon}}]{buironMultiscaleModelMagnetoelastic1999}%
	  \BibitemOpen
	  \bibfield  {author} {\bibinfo {author} {\bibfnamefont {N.}~\bibnamefont
	  {Buiron}}, \bibinfo {author} {\bibfnamefont {L.}~\bibnamefont {Hirsinger}}, \
	  and\ \bibinfo {author} {\bibfnamefont {R.}~\bibnamefont {Billardon}},\ }\href
	  {\doibase 10.1051/jp4:1999919} {\bibfield  {journal} {\bibinfo  {journal} {Le
	  Journal de Physique IV}\ }\textbf {\bibinfo {volume} {09}},\ \bibinfo {pages}
	  {Pr9} (\bibinfo {year} {1999})}\BibitemShut {NoStop}%
	\bibitem [{\citenamefont {Daniel}\ and\ \citenamefont
	  {Galopin}(2008)}]{danielConstitutiveLawMagnetostrictive2008}%
	  \BibitemOpen
	  \bibfield  {author} {\bibinfo {author} {\bibfnamefont {L.}~\bibnamefont
	  {Daniel}}\ and\ \bibinfo {author} {\bibfnamefont {N.}~\bibnamefont
	  {Galopin}},\ }\href {\doibase 10.1051/epjap:2008031} {\bibfield  {journal}
	  {\bibinfo  {journal} {The European Physical Journal Applied Physics}\
	  }\textbf {\bibinfo {volume} {42}},\ \bibinfo {pages} {153} (\bibinfo {year}
	  {2008})}\BibitemShut {NoStop}%
	\bibitem [{\citenamefont {Mei}\ and\ \citenamefont
	  {Vernescu}(2010)}]{meiHomogenizationMethodsMultiscale2010}%
	  \BibitemOpen
	  \bibfield  {author} {\bibinfo {author} {\bibfnamefont {C.~C.}\ \bibnamefont
	  {Mei}}\ and\ \bibinfo {author} {\bibfnamefont {B.}~\bibnamefont {Vernescu}},\
	  }\href@noop {} {{\selectlanguage {English}\emph {\bibinfo {title}
	  {Homogenization Methods for Multiscale Mechanics}}}}\ (\bibinfo  {publisher}
	  {{World Scientific}},\ \bibinfo {address} {{Singapore}},\ \bibinfo {year}
	  {2010})\BibitemShut {NoStop}%
	\bibitem [{\citenamefont
	  {Casalbuoni}(1976{\natexlab{a}})}]{casalbuoniQuantizationSystemsAnticommuting1976}%
	  \BibitemOpen
	  \bibfield  {author} {\bibinfo {author} {\bibfnamefont {R.}~\bibnamefont
	  {Casalbuoni}},\ }\href {\doibase 10.1007/BF02748689} {\bibfield  {journal}
	  {\bibinfo  {journal} {Il Nuovo Cimento A (1965-1970)}\ }\textbf {\bibinfo
	  {volume} {33}},\ \bibinfo {pages} {115} (\bibinfo {year}
	  {1976}{\natexlab{a}})}\BibitemShut {NoStop}%
	\bibitem [{\citenamefont
	  {Casalbuoni}(1976{\natexlab{b}})}]{casalbuoniClassicalMechanicsBosefermi1976}%
	  \BibitemOpen
	  \bibfield  {author} {\bibinfo {author} {\bibfnamefont {R.}~\bibnamefont
	  {Casalbuoni}},\ }\href {\doibase 10.1007/BF02729860} {\bibfield  {journal}
	  {\bibinfo  {journal} {Il Nuovo Cimento A (1965-1970)}\ }\textbf {\bibinfo
	  {volume} {33}},\ \bibinfo {pages} {389} (\bibinfo {year}
	  {1976}{\natexlab{b}})}\BibitemShut {NoStop}%
	\bibitem [{\citenamefont {Berezin}\ and\ \citenamefont
	  {Marinov}(1977)}]{berezinParticleSpinDynamics1977}%
	  \BibitemOpen
	  \bibfield  {author} {\bibinfo {author} {\bibfnamefont {F.}~\bibnamefont
	  {Berezin}}\ and\ \bibinfo {author} {\bibfnamefont {M.}~\bibnamefont
	  {Marinov}},\ }\href {\doibase 10.1016/0003-4916(77)90335-9} {\bibfield
	  {journal} {\bibinfo  {journal} {Annals of Physics}\ }\textbf {\bibinfo
	  {volume} {104}},\ \bibinfo {pages} {336} (\bibinfo {year}
	  {1977})}\BibitemShut {NoStop}%
	\bibitem [{\citenamefont {Nelson}\ and\ \citenamefont
	  {Chen}(1994)}]{nelsonLagrangianTreatmentMagnetic1994}%
	  \BibitemOpen
	  \bibfield  {author} {\bibinfo {author} {\bibfnamefont {D.~F.}\ \bibnamefont
	  {Nelson}}\ and\ \bibinfo {author} {\bibfnamefont {B.}~\bibnamefont {Chen}},\
	  }\href {\doibase 10.1103/PhysRevB.50.1023} {\bibfield  {journal} {\bibinfo
	  {journal} {Physical Review B}\ }\textbf {\bibinfo {volume} {50}},\ \bibinfo
	  {pages} {1023} (\bibinfo {year} {1994})}\BibitemShut {NoStop}%
	\bibitem [{\citenamefont {Cheng}\ \emph {et~al.}(2015)\citenamefont {Cheng},
	  \citenamefont {Daniels}, \citenamefont {Zhu},\ and\ \citenamefont
	  {Xiao}}]{chengUltrafastSwitchingAntiferromagnets2015}%
	  \BibitemOpen
	  \bibfield  {author} {\bibinfo {author} {\bibfnamefont {R.}~\bibnamefont
	  {Cheng}}, \bibinfo {author} {\bibfnamefont {M.~W.}\ \bibnamefont {Daniels}},
	  \bibinfo {author} {\bibfnamefont {J.-G.}\ \bibnamefont {Zhu}}, \ and\
	  \bibinfo {author} {\bibfnamefont {D.}~\bibnamefont {Xiao}},\ }\href {\doibase
	  10.1103/PhysRevB.91.064423} {\bibfield  {journal} {\bibinfo  {journal}
	  {Physical Review B}\ }\textbf {\bibinfo {volume} {91}},\ \bibinfo {pages}
	  {064423} (\bibinfo {year} {2015})}\BibitemShut {NoStop}%
	\bibitem [{\citenamefont {Nussle}\ \emph {et~al.}(2019)\citenamefont {Nussle},
	  \citenamefont {Thibaudeau},\ and\ \citenamefont
	  {Nicolis}}]{nussleCouplingMagnetoelasticLagrangians2019}%
	  \BibitemOpen
	  \bibfield  {author} {\bibinfo {author} {\bibfnamefont {T.}~\bibnamefont
	  {Nussle}}, \bibinfo {author} {\bibfnamefont {P.}~\bibnamefont {Thibaudeau}},
	  \ and\ \bibinfo {author} {\bibfnamefont {S.}~\bibnamefont {Nicolis}},\ }\href
	  {\doibase 10.1016/j.jmmm.2018.09.030} {\bibfield  {journal} {\bibinfo
	  {journal} {Journal of Magnetism and Magnetic Materials}\ }\textbf {\bibinfo
	  {volume} {469}},\ \bibinfo {pages} {633} (\bibinfo {year} {2019})},\ \Eprint
	  {http://arxiv.org/abs/1711.08062} {arXiv:1711.08062} \BibitemShut {NoStop}%
	\bibitem [{\citenamefont
	  {Casimir}(1945)}]{casimirOnsagerPrincipleMicroscopic1945}%
	  \BibitemOpen
	  \bibfield  {author} {\bibinfo {author} {\bibfnamefont {H.~B.~G.}\
	  \bibnamefont {Casimir}},\ }\href {\doibase 10.1103/RevModPhys.17.343}
	  {\bibfield  {journal} {\bibinfo  {journal} {Reviews of Modern Physics}\
	  }\textbf {\bibinfo {volume} {17}},\ \bibinfo {pages} {343} (\bibinfo {year}
	  {1945})}\BibitemShut {NoStop}%
	\bibitem [{\citenamefont {Landau}\ \emph {et~al.}(2008)\citenamefont {Landau},
	  \citenamefont {Lifshitz},\ and\ \citenamefont
	  {Landau}}]{landauTheoryElasticity2008}%
	  \BibitemOpen
	  \bibfield  {author} {\bibinfo {author} {\bibfnamefont {L.~D.}\ \bibnamefont
	  {Landau}}, \bibinfo {author} {\bibfnamefont {E.~M.}\ \bibnamefont
	  {Lifshitz}}, \ and\ \bibinfo {author} {\bibfnamefont {L.~D.}\ \bibnamefont
	  {Landau}},\ }\href@noop {} {{\selectlanguage {English}\emph {\bibinfo {title}
	  {Theory of Elasticity}}}},\ \bibinfo {edition} {3rd}\ ed.,\ \bibinfo {series}
	  {Course of Theoretical Physics}\ No.\ \bibinfo {number} {Vol. 7}\ (\bibinfo
	  {publisher} {{Elsevier}},\ \bibinfo {address} {{Amsterdam}},\ \bibinfo {year}
	  {2008})\BibitemShut {NoStop}%
	\bibitem [{\citenamefont {Gairola}(1978)}]{gairolaNonlocalTheoryElastic1978}%
	  \BibitemOpen
	  \bibfield  {author} {\bibinfo {author} {\bibfnamefont {B.~K.~D.}\
	  \bibnamefont {Gairola}},\ }\href {\doibase 10.1002/pssb.2220850221}
	  {\bibfield  {journal} {\bibinfo  {journal} {Physica Status Solidi (b)}\
	  }\textbf {\bibinfo {volume} {85}},\ \bibinfo {pages} {577} (\bibinfo {year}
	  {1978})}\BibitemShut {NoStop}%
	\bibitem [{\citenamefont {Eringen}(2004)}]{eringenNonlocalContinuumField2004}%
	  \BibitemOpen
	  \bibinfo {editor} {\bibfnamefont {A.~C.}\ \bibnamefont {Eringen}},\ ed.,\
	  \href {\doibase 10.1007/b97697} {{\selectlanguage {English}\emph {\bibinfo
	  {title} {Nonlocal {{Continuum Field Theories}}}}}}\ (\bibinfo  {publisher}
	  {{Springer New York}},\ \bibinfo {address} {{New York, NY}},\ \bibinfo {year}
	  {2004})\BibitemShut {NoStop}%
	\bibitem [{\citenamefont {Toupin}\ and\ \citenamefont
	  {Bernstein}(1961)}]{toupinSoundWavesDeformed1961}%
	  \BibitemOpen
	  \bibfield  {author} {\bibinfo {author} {\bibfnamefont {R.~A.}\ \bibnamefont
	  {Toupin}}\ and\ \bibinfo {author} {\bibfnamefont {B.}~\bibnamefont
	  {Bernstein}},\ }\href {\doibase 10.1121/1.1908623} {\bibfield  {journal}
	  {\bibinfo  {journal} {The Journal of the Acoustical Society of America}\
	  }\textbf {\bibinfo {volume} {33}},\ \bibinfo {pages} {216} (\bibinfo {year}
	  {1961})}\BibitemShut {NoStop}%
	\bibitem [{\citenamefont {Callen}\ and\ \citenamefont
	  {Callen}(1963)}]{callenStaticMagnetoelasticCoupling1963}%
	  \BibitemOpen
	  \bibfield  {author} {\bibinfo {author} {\bibfnamefont {E.~R.}\ \bibnamefont
	  {Callen}}\ and\ \bibinfo {author} {\bibfnamefont {H.~B.}\ \bibnamefont
	  {Callen}},\ }\href {\doibase 10.1103/PhysRev.129.578} {\bibfield  {journal}
	  {\bibinfo  {journal} {Physical Review}\ }\textbf {\bibinfo {volume} {129}},\
	  \bibinfo {pages} {578} (\bibinfo {year} {1963})}\BibitemShut {NoStop}%
	\bibitem [{\citenamefont {Brown}(1966)}]{brownMagnetoelasticInteractions1966}%
	  \BibitemOpen
	  \bibfield  {author} {\bibinfo {author} {\bibfnamefont {W.~F.}\ \bibnamefont
	  {Brown}},\ }\href {\doibase 10.1007/978-3-642-87396-6} {\emph {\bibinfo
	  {title} {Magnetoelastic {{Interactions}}}}},\ \bibinfo {edition} {english}\
	  ed.,\ edited by\ \bibinfo {editor} {\bibfnamefont {C.}~\bibnamefont
	  {Truesdell}}, \bibinfo {editor} {\bibfnamefont {R.}~\bibnamefont {Aris}},
	  \bibinfo {editor} {\bibfnamefont {L.}~\bibnamefont {Collatz}}, \bibinfo
	  {editor} {\bibfnamefont {G.}~\bibnamefont {Fichera}}, \bibinfo {editor}
	  {\bibfnamefont {P.}~\bibnamefont {Germain}}, \bibinfo {editor} {\bibfnamefont
	  {J.}~\bibnamefont {Keller}}, \bibinfo {editor} {\bibfnamefont {M.~M.}\
	  \bibnamefont {Schiffer}}, \ and\ \bibinfo {editor} {\bibfnamefont
	  {A.}~\bibnamefont {Seeger}},\ \bibinfo {series} {Springer {{Tracts}} in
	  {{Natural Philosophy}}}, Vol.~\bibinfo {volume} {9}\ (\bibinfo  {publisher}
	  {{Springer Berlin Heidelberg}},\ \bibinfo {address} {{Berlin, Heidelberg}},\
	  \bibinfo {year} {1966})\BibitemShut {NoStop}%
	\bibitem [{\citenamefont {Rosen}(1972)}]{rosenGalileanInvarianceGeneral1972}%
	  \BibitemOpen
	  \bibfield  {author} {\bibinfo {author} {\bibfnamefont {G.}~\bibnamefont
	  {Rosen}},\ }\href {\doibase 10.1119/1.1986618} {\bibfield  {journal}
	  {\bibinfo  {journal} {American Journal of Physics}\ }\textbf {\bibinfo
	  {volume} {40}},\ \bibinfo {pages} {683} (\bibinfo {year} {1972})}\BibitemShut
	  {NoStop}%
	\bibitem [{\citenamefont
	  {Tiersten}(1964)}]{tierstenCoupledMagnetomechanicalEquations1964}%
	  \BibitemOpen
	  \bibfield  {author} {\bibinfo {author} {\bibfnamefont {H.~F.}\ \bibnamefont
	  {Tiersten}},\ }\href {\doibase 10.1063/1.1704239} {\bibfield  {journal}
	  {\bibinfo  {journal} {Journal of Mathematical Physics}\ }\textbf {\bibinfo
	  {volume} {5}},\ \bibinfo {pages} {1298} (\bibinfo {year} {1964})}\BibitemShut
	  {NoStop}%
	\bibitem [{\citenamefont {Akhiezer}\ \emph {et~al.}(1968)\citenamefont
	  {Akhiezer}, \citenamefont {Bar'yakhtar},\ and\ \citenamefont
	  {Peletminskii}}]{akhiezerSpinWaves1968}%
	  \BibitemOpen
	  \bibfield  {author} {\bibinfo {author} {\bibfnamefont {A.}~\bibnamefont
	  {Akhiezer}}, \bibinfo {author} {\bibfnamefont {V.}~\bibnamefont
	  {Bar'yakhtar}}, \ and\ \bibinfo {author} {\bibfnamefont {S.}~\bibnamefont
	  {Peletminskii}},\ }\href@noop {} {{\selectlanguage {English}\emph {\bibinfo
	  {title} {Spin {{Waves}}}}}},\ \bibinfo {series} {North {{Holland}} Series in
	  Low Temperature Physics}, Vol.~\bibinfo {volume} {1}\ (\bibinfo {address}
	  {{Amsterdam}},\ \bibinfo {year} {1968})\BibitemShut {NoStop}%
	\bibitem [{\citenamefont
	  {Dever}(1972)}]{deverTemperatureDependenceElastic1972}%
	  \BibitemOpen
	  \bibfield  {author} {\bibinfo {author} {\bibfnamefont {D.}~\bibnamefont
	  {Dever}},\ }\href {\doibase 10.1063/1.1661710} {\bibfield  {journal}
	  {\bibinfo  {journal} {Journal of Applied Physics}\ }\textbf {\bibinfo
	  {volume} {43}},\ \bibinfo {pages} {3293} (\bibinfo {year}
	  {1972})}\BibitemShut {NoStop}%
	\bibitem [{\citenamefont {Wilczek}(2009)}]{wilczekMajoranaReturns2009}%
	  \BibitemOpen
	  \bibfield  {author} {\bibinfo {author} {\bibfnamefont {F.}~\bibnamefont
	  {Wilczek}},\ }\href {\doibase 10.1038/nphys1380} {\bibfield  {journal}
	  {\bibinfo  {journal} {Nature Physics}\ }\textbf {\bibinfo {volume} {5}},\
	  \bibinfo {pages} {614} (\bibinfo {year} {2009})}\BibitemShut {NoStop}%
	\bibitem [{\citenamefont {Alicea}(2012)}]{aliceaNewDirectionsPursuit2012}%
	  \BibitemOpen
	  \bibfield  {author} {\bibinfo {author} {\bibfnamefont {J.}~\bibnamefont
	  {Alicea}},\ }\href {\doibase 10.1088/0034-4885/75/7/076501} {\bibfield
	  {journal} {\bibinfo  {journal} {Reports on Progress in Physics}\ }\textbf
	  {\bibinfo {volume} {75}},\ \bibinfo {pages} {076501} (\bibinfo {year}
	  {2012})}\BibitemShut {NoStop}%
	\bibitem [{\citenamefont
	  {Lounesto}(2003)}]{lounestoCliffordAlgebrasSpinors2003}%
	  \BibitemOpen
	  \bibfield  {author} {\bibinfo {author} {\bibfnamefont {P.}~\bibnamefont
	  {Lounesto}},\ }\href@noop {} {{\selectlanguage {English}\emph {\bibinfo
	  {title} {Clifford Algebras and Spinors}}}},\ \bibinfo {edition} {2nd}\ ed.,\
	  \bibinfo {series} {London {{Mathematical Society}} Lecture Note Series}\ No.\
	  \bibinfo {number} {286}\ (\bibinfo  {publisher} {{Cambridge Univ. Press}},\
	  \bibinfo {address} {{Cambridge}},\ \bibinfo {year} {2003})\BibitemShut
	  {NoStop}%
	\bibitem [{\citenamefont {Varshalovich}\ \emph {et~al.}(1988)\citenamefont
	  {Varshalovich}, \citenamefont {Moskalev},\ and\ \citenamefont
	  {Khersonskii}}]{varshalovichQuantumTheoryAngular1988}%
	  \BibitemOpen
	  \bibfield  {author} {\bibinfo {author} {\bibfnamefont {D.~A.}\ \bibnamefont
	  {Varshalovich}}, \bibinfo {author} {\bibfnamefont {A.~N.}\ \bibnamefont
	  {Moskalev}}, \ and\ \bibinfo {author} {\bibfnamefont {V.~K.}\ \bibnamefont
	  {Khersonskii}},\ }\href {\doibase 10.1142/0270} {{\selectlanguage
	  {English}\emph {\bibinfo {title} {Quantum {{Theory}} of {{Angular
	  Momentum}}}}}}\ (\bibinfo  {publisher} {{World Scientific}},\ \bibinfo {year}
	  {1988})\BibitemShut {NoStop}%
	\bibitem [{\citenamefont {Evans}\ \emph {et~al.}(2014)\citenamefont {Evans},
	  \citenamefont {Fan}, \citenamefont {Chureemart}, \citenamefont {Ostler},
	  \citenamefont {Ellis},\ and\ \citenamefont
	  {Chantrell}}]{evansAtomisticSpinModel2014}%
	  \BibitemOpen
	  \bibfield  {author} {\bibinfo {author} {\bibfnamefont {R.~F.}\ \bibnamefont
	  {Evans}}, \bibinfo {author} {\bibfnamefont {W.~J.}\ \bibnamefont {Fan}},
	  \bibinfo {author} {\bibfnamefont {P.}~\bibnamefont {Chureemart}}, \bibinfo
	  {author} {\bibfnamefont {T.~A.}\ \bibnamefont {Ostler}}, \bibinfo {author}
	  {\bibfnamefont {M.~O.}\ \bibnamefont {Ellis}}, \ and\ \bibinfo {author}
	  {\bibfnamefont {R.~W.}\ \bibnamefont {Chantrell}},\ }\href {\doibase
	  10.1088/0953-8984/26/10/103202} {\bibfield  {journal} {\bibinfo  {journal}
	  {Journal of Physics: Condensed Matter}\ }\textbf {\bibinfo {volume} {26}},\
	  \bibinfo {pages} {103202} (\bibinfo {year} {2014})}\BibitemShut {NoStop}%
	\bibitem [{\citenamefont {Eriksson}\ \emph {et~al.}(2017)\citenamefont
	  {Eriksson}, \citenamefont {Bergman}, \citenamefont {Bergqvist},\ and\
	  \citenamefont {Hellsvik}}]{erikssonAtomisticSpinDynamics2017}%
	  \BibitemOpen
	  \bibfield  {author} {\bibinfo {author} {\bibfnamefont {O.}~\bibnamefont
	  {Eriksson}}, \bibinfo {author} {\bibfnamefont {A.}~\bibnamefont {Bergman}},
	  \bibinfo {author} {\bibfnamefont {L.}~\bibnamefont {Bergqvist}}, \ and\
	  \bibinfo {author} {\bibfnamefont {J.}~\bibnamefont {Hellsvik}},\ }\href@noop
	  {} {{\selectlanguage {English}\emph {\bibinfo {title} {Atomistic Spin
	  Dynamics: Foundations and Applications}}}},\ \bibinfo {edition} {first
	  edition}\ ed.\ (\bibinfo  {publisher} {{Oxford University Press}},\ \bibinfo
	  {address} {{Oxford}},\ \bibinfo {year} {2017})\BibitemShut {NoStop}%
	\bibitem [{\citenamefont {{van
	  Vleck}}(1937)}]{vanvleckAnisotropyCubicFerromagnetic1937}%
	  \BibitemOpen
	  \bibfield  {author} {\bibinfo {author} {\bibfnamefont {J.~H.}\ \bibnamefont
	  {{van Vleck}}},\ }\href {\doibase 10.1103/PhysRev.52.1178} {\bibfield
	  {journal} {\bibinfo  {journal} {Physical Review}\ }\textbf {\bibinfo {volume}
	  {52}},\ \bibinfo {pages} {1178} (\bibinfo {year} {1937})}\BibitemShut
	  {NoStop}%
	\bibitem [{\citenamefont
	  {N{\'e}el}(1954)}]{neelAnisotropieMagnetiqueSuperficielle1954}%
	  \BibitemOpen
	  \bibfield  {author} {\bibinfo {author} {\bibfnamefont {L.}~\bibnamefont
	  {N{\'e}el}},\ }\href {\doibase 10.1051/jphysrad:01954001504022500} {\bibfield
	   {journal} {\bibinfo  {journal} {Journal de Physique et le Radium}\ }\textbf
	  {\bibinfo {volume} {15}},\ \bibinfo {pages} {225} (\bibinfo {year}
	  {1954})}\BibitemShut {NoStop}%
	\bibitem [{\citenamefont
	  {Lee}(1955)}]{leeMagnetostrictionMagnetomechanicalEffects1955}%
	  \BibitemOpen
	  \bibfield  {author} {\bibinfo {author} {\bibfnamefont {E.~W.}\ \bibnamefont
	  {Lee}},\ }\href {\doibase 10.1088/0034-4885/18/1/305} {\bibfield  {journal}
	  {\bibinfo  {journal} {Reports on Progress in Physics}\ }\textbf {\bibinfo
	  {volume} {18}},\ \bibinfo {pages} {184} (\bibinfo {year} {1955})}\BibitemShut
	  {NoStop}%
	\bibitem [{\citenamefont {Sabiryanov}\ and\ \citenamefont
	  {Jaswal}(1999)}]{sabiryanovMagnonsMagnonPhononInteractions1999}%
	  \BibitemOpen
	  \bibfield  {author} {\bibinfo {author} {\bibfnamefont {R.~F.}\ \bibnamefont
	  {Sabiryanov}}\ and\ \bibinfo {author} {\bibfnamefont {S.~S.}\ \bibnamefont
	  {Jaswal}},\ }\href {\doibase 10.1103/PhysRevLett.83.2062} {\bibfield
	  {journal} {\bibinfo  {journal} {Physical Review Letters}\ }\textbf {\bibinfo
	  {volume} {83}},\ \bibinfo {pages} {2062} (\bibinfo {year}
	  {1999})}\BibitemShut {NoStop}%
	\bibitem [{\citenamefont
	  {Kastrup}(1983)}]{kastrupCanonicalTheoriesLagrangian1983}%
	  \BibitemOpen
	  \bibfield  {author} {\bibinfo {author} {\bibfnamefont {H.~A.}\ \bibnamefont
	  {Kastrup}},\ }\href {\doibase 10.1016/0370-1573(83)90037-6} {\bibfield
	  {journal} {\bibinfo  {journal} {Physics Reports}\ }\textbf {\bibinfo {volume}
	  {101}},\ \bibinfo {pages} {1} (\bibinfo {year} {1983})}\BibitemShut {NoStop}%
	\bibitem [{\citenamefont
	  {Kanatchikov}(1993)}]{kanatchikovCanonicalStructureDonderWeyl1993}%
	  \BibitemOpen
	  \bibfield  {author} {\bibinfo {author} {\bibfnamefont {I.~V.}\ \bibnamefont
	  {Kanatchikov}},\ }\href@noop {} {\bibfield  {journal} {\bibinfo  {journal}
	  {arXiv:9312162 (hep-th)}\ } (\bibinfo {year} {1993})}\BibitemShut {NoStop}%
	\bibitem [{\citenamefont {Arnowitt}\ \emph {et~al.}(1960)\citenamefont
	  {Arnowitt}, \citenamefont {Deser},\ and\ \citenamefont
	  {Misner}}]{arnowittCanonicalVariablesGeneral1960}%
	  \BibitemOpen
	  \bibfield  {author} {\bibinfo {author} {\bibfnamefont {R.}~\bibnamefont
	  {Arnowitt}}, \bibinfo {author} {\bibfnamefont {S.}~\bibnamefont {Deser}}, \
	  and\ \bibinfo {author} {\bibfnamefont {C.~W.}\ \bibnamefont {Misner}},\
	  }\href {\doibase 10.1103/PhysRev.117.1595} {\bibfield  {journal} {\bibinfo
	  {journal} {Physical Review}\ }\textbf {\bibinfo {volume} {117}},\ \bibinfo
	  {pages} {1595} (\bibinfo {year} {1960})}\BibitemShut {NoStop}%
	\bibitem [{\citenamefont {Hojman}\ \emph {et~al.}(1976)\citenamefont {Hojman},
	  \citenamefont {Kucha{\v r}},\ and\ \citenamefont
	  {Teitelboim}}]{hojmanGeometrodynamicsRegained1976}%
	  \BibitemOpen
	  \bibfield  {author} {\bibinfo {author} {\bibfnamefont {S.~A.}\ \bibnamefont
	  {Hojman}}, \bibinfo {author} {\bibfnamefont {K.}~\bibnamefont {Kucha{\v r}}},
	  \ and\ \bibinfo {author} {\bibfnamefont {C.}~\bibnamefont {Teitelboim}},\
	  }\href {\doibase 10.1016/0003-4916(76)90112-3} {\bibfield  {journal}
	  {\bibinfo  {journal} {Annals of Physics}\ }\textbf {\bibinfo {volume} {96}},\
	  \bibinfo {pages} {88} (\bibinfo {year} {1976})}\BibitemShut {NoStop}%
	\bibitem [{\citenamefont
	  {Hill}(1970)}]{hillConstitutiveInequalitiesIsotropic1970}%
	  \BibitemOpen
	  \bibfield  {author} {\bibinfo {author} {\bibfnamefont {R.}~\bibnamefont
	  {Hill}},\ }\href {\doibase 10.1098/rspa.1970.0018} {\bibfield  {journal}
	  {\bibinfo  {journal} {Proceedings of the Royal Society of London. A.
	  Mathematical and Physical Sciences}\ }\textbf {\bibinfo {volume} {314}},\
	  \bibinfo {pages} {457} (\bibinfo {year} {1970})}\BibitemShut {NoStop}%
	\bibitem [{\citenamefont {Kijowski}\ and\ \citenamefont
	  {Szczyrba}(1976)}]{kijowskiCanonicalStructureClassical1976}%
	  \BibitemOpen
	  \bibfield  {author} {\bibinfo {author} {\bibfnamefont {J.}~\bibnamefont
	  {Kijowski}}\ and\ \bibinfo {author} {\bibfnamefont {W.}~\bibnamefont
	  {Szczyrba}},\ }\href {\doibase 10.1007/BF01608496} {\bibfield  {journal}
	  {\bibinfo  {journal} {Communications in Mathematical Physics}\ }\textbf
	  {\bibinfo {volume} {46}},\ \bibinfo {pages} {183} (\bibinfo {year}
	  {1976})}\BibitemShut {NoStop}%
	\bibitem [{\citenamefont
	  {Szczyrba}(1976)}]{szczyrbaSymplecticStructureSet1976}%
	  \BibitemOpen
	  \bibfield  {author} {\bibinfo {author} {\bibfnamefont {W.}~\bibnamefont
	  {Szczyrba}},\ }\href {\doibase 10.1007/BF01609347} {\bibfield  {journal}
	  {\bibinfo  {journal} {Communications in Mathematical Physics}\ }\textbf
	  {\bibinfo {volume} {51}},\ \bibinfo {pages} {163} (\bibinfo {year}
	  {1976})}\BibitemShut {NoStop}%
	\bibitem [{\citenamefont {Marsden}\ \emph {et~al.}(1986)\citenamefont
	  {Marsden}, \citenamefont {Montgomery}, \citenamefont {Morrison},\ and\
	  \citenamefont {Thompson}}]{marsdenCovariantPoissonBrackets1986}%
	  \BibitemOpen
	  \bibfield  {author} {\bibinfo {author} {\bibfnamefont {J.~E.}\ \bibnamefont
	  {Marsden}}, \bibinfo {author} {\bibfnamefont {R.}~\bibnamefont {Montgomery}},
	  \bibinfo {author} {\bibfnamefont {P.~J.}\ \bibnamefont {Morrison}}, \ and\
	  \bibinfo {author} {\bibfnamefont {W.~B.}\ \bibnamefont {Thompson}},\ }\href
	  {\doibase 10.1016/0003-4916(86)90157-0} {\bibfield  {journal} {\bibinfo
	  {journal} {Annals of Physics}\ }\textbf {\bibinfo {volume} {169}},\ \bibinfo
	  {pages} {29} (\bibinfo {year} {1986})}\BibitemShut {NoStop}%
	\bibitem [{\citenamefont {Yang}\ and\ \citenamefont
	  {Hirschfelder}(1980)}]{yangGeneralizationsClassicalPoisson1980}%
	  \BibitemOpen
	  \bibfield  {author} {\bibinfo {author} {\bibfnamefont {K.-H.}\ \bibnamefont
	  {Yang}}\ and\ \bibinfo {author} {\bibfnamefont {J.~O.}\ \bibnamefont
	  {Hirschfelder}},\ }\href {\doibase 10.1103/PhysRevA.22.1814} {\bibfield
	  {journal} {\bibinfo  {journal} {Physical Review A}\ }\textbf {\bibinfo
	  {volume} {22}},\ \bibinfo {pages} {1814} (\bibinfo {year}
	  {1980})}\BibitemShut {NoStop}%
	\bibitem [{\citenamefont {Skomski}(2008)}]{skomskiSimpleModelsMagnetism2008}%
	  \BibitemOpen
	  \bibfield  {author} {\bibinfo {author} {\bibfnamefont {R.}~\bibnamefont
	  {Skomski}},\ }\href@noop {} {{\selectlanguage {English}\emph {\bibinfo
	  {title} {Simple {{Models}} of {{Magnetism}}}}}},\ Oxford {{Graduate Texts}}\
	  (\bibinfo  {publisher} {{Oxford University Press}},\ \bibinfo {address}
	  {{Oxford, New York}},\ \bibinfo {year} {2008})\BibitemShut {NoStop}%
	\bibitem [{\citenamefont {Blanes}\ \emph {et~al.}(2009)\citenamefont {Blanes},
	  \citenamefont {Casas}, \citenamefont {Oteo},\ and\ \citenamefont
	  {Ros}}]{blanesMagnusExpansionIts2009}%
	  \BibitemOpen
	  \bibfield  {author} {\bibinfo {author} {\bibfnamefont {S.}~\bibnamefont
	  {Blanes}}, \bibinfo {author} {\bibfnamefont {F.}~\bibnamefont {Casas}},
	  \bibinfo {author} {\bibfnamefont {J.~A.}\ \bibnamefont {Oteo}}, \ and\
	  \bibinfo {author} {\bibfnamefont {J.}~\bibnamefont {Ros}},\ }\href {\doibase
	  10.1016/j.physrep.2008.11.001} {\bibfield  {journal} {\bibinfo  {journal}
	  {Physics Reports}\ }\textbf {\bibinfo {volume} {470}},\ \bibinfo {pages}
	  {151} (\bibinfo {year} {2009})}\BibitemShut {NoStop}%
	\bibitem [{\citenamefont {Hairer}\ \emph {et~al.}(2006)\citenamefont {Hairer},
	  \citenamefont {Lubich},\ and\ \citenamefont
	  {Wanner}}]{hairerGeometricNumericalIntegration2006}%
	  \BibitemOpen
	  \bibfield  {author} {\bibinfo {author} {\bibfnamefont {E.}~\bibnamefont
	  {Hairer}}, \bibinfo {author} {\bibfnamefont {C.}~\bibnamefont {Lubich}}, \
	  and\ \bibinfo {author} {\bibfnamefont {G.}~\bibnamefont {Wanner}},\
	  }\href@noop {} {{\selectlanguage {English}\emph {\bibinfo {title} {Geometric
	  {{Numerical Integration}}: {{Structure}}-{{Preserving Algorithms}} for
	  {{Ordinary Differential Equations}}}}}},\ \bibinfo {edition} {2nd}\ ed.,\
	  Springer {{Series}} in {{Computational Mathematics}}\ (\bibinfo  {publisher}
	  {{Springer-Verlag}},\ \bibinfo {address} {{Berlin Heidelberg}},\ \bibinfo
	  {year} {2006})\BibitemShut {NoStop}%
	\bibitem [{\citenamefont {Batcho}\ and\ \citenamefont
	  {Schlick}(2001)}]{batchoSpecialStabilityAdvantages2001}%
	  \BibitemOpen
	  \bibfield  {author} {\bibinfo {author} {\bibfnamefont {P.~F.}\ \bibnamefont
	  {Batcho}}\ and\ \bibinfo {author} {\bibfnamefont {T.}~\bibnamefont
	  {Schlick}},\ }\href {\doibase 10.1063/1.1389855} {\bibfield  {journal}
	  {\bibinfo  {journal} {The Journal of Chemical Physics}\ }\textbf {\bibinfo
	  {volume} {115}},\ \bibinfo {pages} {4019} (\bibinfo {year}
	  {2001})}\BibitemShut {NoStop}%
	\bibitem [{\citenamefont {Beaujouan}\ \emph {et~al.}(2012)\citenamefont
	  {Beaujouan}, \citenamefont {Thibaudeau},\ and\ \citenamefont
	  {Barreteau}}]{beaujouanAnisotropicMagneticMolecular2012}%
	  \BibitemOpen
	  \bibfield  {author} {\bibinfo {author} {\bibfnamefont {D.}~\bibnamefont
	  {Beaujouan}}, \bibinfo {author} {\bibfnamefont {P.}~\bibnamefont
	  {Thibaudeau}}, \ and\ \bibinfo {author} {\bibfnamefont {C.}~\bibnamefont
	  {Barreteau}},\ }\href {\doibase 10.1103/PhysRevB.86.174409} {\bibfield
	  {journal} {\bibinfo  {journal} {Physical Review B}\ }\textbf {\bibinfo
	  {volume} {86}},\ \bibinfo {pages} {174409} (\bibinfo {year}
	  {2012})}\BibitemShut {NoStop}%
	\bibitem [{\citenamefont {Rodrigues}(1840)}]{rodriguesLoisGeometriquesQui1840}%
	  \BibitemOpen
	  \bibfield  {author} {\bibinfo {author} {\bibfnamefont {O.}~\bibnamefont
	  {Rodrigues}},\ }\href@noop {} {\bibfield  {journal} {\bibinfo  {journal}
	  {Journal de Math{\'e}matiques Pures et Appliqu{\'e}es}\ }\textbf {\bibinfo
	  {volume} {5}},\ \bibinfo {pages} {380} (\bibinfo {year} {1840})}\BibitemShut
	  {NoStop}%
	\bibitem [{\citenamefont {Honerkamp}\ and\ \citenamefont
	  {R{\"o}mer}(1993)}]{honerkampTheoreticalPhysicsClassical1993}%
	  \BibitemOpen
	  \bibfield  {author} {\bibinfo {author} {\bibfnamefont {J.}~\bibnamefont
	  {Honerkamp}}\ and\ \bibinfo {author} {\bibfnamefont {H.}~\bibnamefont
	  {R{\"o}mer}},\ }\href@noop {} {{\selectlanguage {English}\emph {\bibinfo
	  {title} {Theoretical Physics: A Classical Approach}}}}\ (\bibinfo
	  {publisher} {{Springer}},\ \bibinfo {address} {{Berlin; New York}},\ \bibinfo
	  {year} {1993})\BibitemShut {NoStop}%
	\bibitem [{\citenamefont {Thibaudeau}\ and\ \citenamefont
	  {Beaujouan}(2012)}]{thibaudeauThermostattingAtomicSpin2012}%
	  \BibitemOpen
	  \bibfield  {author} {\bibinfo {author} {\bibfnamefont {P.}~\bibnamefont
	  {Thibaudeau}}\ and\ \bibinfo {author} {\bibfnamefont {D.}~\bibnamefont
	  {Beaujouan}},\ }\href {\doibase 10.1016/j.physa.2011.11.030} {\bibfield
	  {journal} {\bibinfo  {journal} {Physica A: Statistical Mechanics and its
	  Applications}\ }\textbf {\bibinfo {volume} {391}},\ \bibinfo {pages} {1963}
	  (\bibinfo {year} {2012})}\BibitemShut {NoStop}%
	\bibitem [{\citenamefont {Omelyan}\ \emph {et~al.}(2003)\citenamefont
	  {Omelyan}, \citenamefont {Mryglod},\ and\ \citenamefont
	  {Folk}}]{omelyanSymplecticAnalyticallyIntegrable2003}%
	  \BibitemOpen
	  \bibfield  {author} {\bibinfo {author} {\bibfnamefont {I.~P.}\ \bibnamefont
	  {Omelyan}}, \bibinfo {author} {\bibfnamefont {I.~M.}\ \bibnamefont
	  {Mryglod}}, \ and\ \bibinfo {author} {\bibfnamefont {R.}~\bibnamefont
	  {Folk}},\ }\href {\doibase 10.1016/S0010-4655(02)00754-3} {\bibfield
	  {journal} {\bibinfo  {journal} {Computer Physics Communications}\ }\textbf
	  {\bibinfo {volume} {151}},\ \bibinfo {pages} {272} (\bibinfo {year}
	  {2003})}\BibitemShut {NoStop}%
	\bibitem [{\citenamefont
	  {Ogden}(1997)}]{ogdenNonlinearElasticDeformations1997}%
	  \BibitemOpen
	  \bibfield  {author} {\bibinfo {author} {\bibfnamefont {R.~W.}\ \bibnamefont
	  {Ogden}},\ }\href@noop {} {{\selectlanguage {English}\emph {\bibinfo {title}
	  {Non-Linear Elastic Deformations}}}},\ \bibinfo {edition} {dover ed}\ ed.,\
	  Dover Books on Physics\ (\bibinfo  {publisher} {{Dover}},\ \bibinfo {address}
	  {{Mineola, NY}},\ \bibinfo {year} {1997})\BibitemShut {NoStop}%
	\bibitem [{\citenamefont {F{\"a}hnle}\ \emph {et~al.}(2002)\citenamefont
	  {F{\"a}hnle}, \citenamefont {Komelj}, \citenamefont {Wu},\ and\ \citenamefont
	  {Guo}}]{fahnleMagnetoelasticityFePossible2002}%
	  \BibitemOpen
	  \bibfield  {author} {\bibinfo {author} {\bibfnamefont {M.}~\bibnamefont
	  {F{\"a}hnle}}, \bibinfo {author} {\bibfnamefont {M.}~\bibnamefont {Komelj}},
	  \bibinfo {author} {\bibfnamefont {R.~Q.}\ \bibnamefont {Wu}}, \ and\ \bibinfo
	  {author} {\bibfnamefont {G.~Y.}\ \bibnamefont {Guo}},\ }\href {\doibase
	  10.1103/PhysRevB.65.144436} {\bibfield  {journal} {\bibinfo  {journal}
	  {Physical Review B}\ }\textbf {\bibinfo {volume} {65}},\ \bibinfo {pages}
	  {144436} (\bibinfo {year} {2002})}\BibitemShut {NoStop}%
	\bibitem [{\citenamefont {Tranchida}\ \emph {et~al.}(2018)\citenamefont
	  {Tranchida}, \citenamefont {Thibaudeau},\ and\ \citenamefont
	  {Nicolis}}]{tranchidaHierarchiesLandauLifshitzBlochEquations2018}%
	  \BibitemOpen
	  \bibfield  {author} {\bibinfo {author} {\bibfnamefont {J.}~\bibnamefont
	  {Tranchida}}, \bibinfo {author} {\bibfnamefont {P.}~\bibnamefont
	  {Thibaudeau}}, \ and\ \bibinfo {author} {\bibfnamefont {S.}~\bibnamefont
	  {Nicolis}},\ }\href {\doibase 10.1103/PhysRevE.98.042101} {\bibfield
	  {journal} {\bibinfo  {journal} {Physical Review E}\ }\textbf {\bibinfo
	  {volume} {98}},\ \bibinfo {pages} {042101} (\bibinfo {year}
	  {2018})}\BibitemShut {NoStop}%
	\bibitem [{\citenamefont {Brink}\ \emph {et~al.}(1976)\citenamefont {Brink},
	  \citenamefont {Di~Vecchia},\ and\ \citenamefont
	  {Howe}}]{brinkLocallySupersymmetricReparametrization1976}%
	  \BibitemOpen
	  \bibfield  {author} {\bibinfo {author} {\bibfnamefont {L.}~\bibnamefont
	  {Brink}}, \bibinfo {author} {\bibfnamefont {P.}~\bibnamefont {Di~Vecchia}}, \
	  and\ \bibinfo {author} {\bibfnamefont {P.}~\bibnamefont {Howe}},\ }\href
	  {\doibase 10.1016/0370-2693(76)90445-7} {\bibfield  {journal} {\bibinfo
	  {journal} {Physics Letters B}\ }\textbf {\bibinfo {volume} {65}},\ \bibinfo
	  {pages} {471} (\bibinfo {year} {1976})}\BibitemShut {NoStop}%
	\bibitem [{\citenamefont {Brink}\ \emph {et~al.}(1977)\citenamefont {Brink},
	  \citenamefont {Di~Vecchia},\ and\ \citenamefont
	  {Howe}}]{brinkLagrangianFormulationClassical1977}%
	  \BibitemOpen
	  \bibfield  {author} {\bibinfo {author} {\bibfnamefont {L.}~\bibnamefont
	  {Brink}}, \bibinfo {author} {\bibfnamefont {P.}~\bibnamefont {Di~Vecchia}}, \
	  and\ \bibinfo {author} {\bibfnamefont {P.}~\bibnamefont {Howe}},\ }\href
	  {\doibase 10.1016/0550-3213(77)90364-9} {\bibfield  {journal} {\bibinfo
	  {journal} {Nuclear Physics B}\ }\textbf {\bibinfo {volume} {118}},\ \bibinfo
	  {pages} {76} (\bibinfo {year} {1977})}\BibitemShut {NoStop}%
	\bibitem [{\citenamefont {Gomis}\ \emph {et~al.}(1995)\citenamefont {Gomis},
	  \citenamefont {Par{\'i}s},\ and\ \citenamefont
	  {Samuel}}]{gomisAntibracketAntifieldsGaugetheory1995}%
	  \BibitemOpen
	  \bibfield  {author} {\bibinfo {author} {\bibfnamefont {J.}~\bibnamefont
	  {Gomis}}, \bibinfo {author} {\bibfnamefont {J.}~\bibnamefont {Par{\'i}s}}, \
	  and\ \bibinfo {author} {\bibfnamefont {S.}~\bibnamefont {Samuel}},\ }\href
	  {\doibase 10.1016/0370-1573(94)00112-G} {\bibfield  {journal} {\bibinfo
	  {journal} {Physics Reports}\ }\textbf {\bibinfo {volume} {259}},\ \bibinfo
	  {pages} {1} (\bibinfo {year} {1995})}\BibitemShut {NoStop}%
	\bibitem [{\citenamefont
	  {Kanatchikov}(1997)}]{kanatchikovFieldTheoreticGeneralizations1997}%
	  \BibitemOpen
	  \bibfield  {author} {\bibinfo {author} {\bibfnamefont {I.~V.}\ \bibnamefont
	  {Kanatchikov}},\ }\href {\doibase 10.1016/S0034-4877(97)85919-8} {\bibfield
	  {journal} {\bibinfo  {journal} {Reports on Mathematical Physics}\ }\textbf
	  {\bibinfo {volume} {40}},\ \bibinfo {pages} {225} (\bibinfo {year}
	  {1997})}\BibitemShut {NoStop}%
	\bibitem [{\citenamefont
	  {Casalbuoni}(1976{\natexlab{c}})}]{casalbuoniRelativelySupersymmetries1976}%
	  \BibitemOpen
	  \bibfield  {author} {\bibinfo {author} {\bibfnamefont {R.}~\bibnamefont
	  {Casalbuoni}},\ }\href {\doibase 10.1016/0370-2693(76)90044-7} {\bibfield
	  {journal} {\bibinfo  {journal} {Physics Letters B}\ }\textbf {\bibinfo
	  {volume} {62}},\ \bibinfo {pages} {49} (\bibinfo {year}
	  {1976}{\natexlab{c}})}\BibitemShut {NoStop}%
	\bibitem [{\citenamefont {Shnirman}\ and\ \citenamefont
	  {Makhlin}(2003)}]{shnirmanSpinSpinCorrelatorsMajorana2003}%
	  \BibitemOpen
	  \bibfield  {author} {\bibinfo {author} {\bibfnamefont {A.}~\bibnamefont
	  {Shnirman}}\ and\ \bibinfo {author} {\bibfnamefont {Y.}~\bibnamefont
	  {Makhlin}},\ }\href {\doibase 10.1103/PhysRevLett.91.207204} {\bibfield
	  {journal} {\bibinfo  {journal} {Physical Review Letters}\ }\textbf {\bibinfo
	  {volume} {91}},\ \bibinfo {pages} {207204} (\bibinfo {year}
	  {2003})}\BibitemShut {NoStop}%
	\bibitem [{\citenamefont {Mao}\ \emph {et~al.}(2003)\citenamefont {Mao},
	  \citenamefont {Coleman}, \citenamefont {Hooley},\ and\ \citenamefont
	  {Langreth}}]{maoSpinDynamicsMajorana2003}%
	  \BibitemOpen
	  \bibfield  {author} {\bibinfo {author} {\bibfnamefont {W.}~\bibnamefont
	  {Mao}}, \bibinfo {author} {\bibfnamefont {P.}~\bibnamefont {Coleman}},
	  \bibinfo {author} {\bibfnamefont {C.}~\bibnamefont {Hooley}}, \ and\ \bibinfo
	  {author} {\bibfnamefont {D.}~\bibnamefont {Langreth}},\ }\href {\doibase
	  10.1103/PhysRevLett.91.207203} {\bibfield  {journal} {\bibinfo  {journal}
	  {Physical Review Letters}\ }\textbf {\bibinfo {volume} {91}},\ \bibinfo
	  {pages} {207203} (\bibinfo {year} {2003})}\BibitemShut {NoStop}%
	\bibitem [{\citenamefont {Schad}\ \emph {et~al.}(2015)\citenamefont {Schad},
	  \citenamefont {Makhlin}, \citenamefont {Narozhny}, \citenamefont
	  {Sch{\"o}n},\ and\ \citenamefont
	  {Shnirman}}]{schadMajoranaRepresentationDissipative2015}%
	  \BibitemOpen
	  \bibfield  {author} {\bibinfo {author} {\bibfnamefont {P.}~\bibnamefont
	  {Schad}}, \bibinfo {author} {\bibfnamefont {Y.}~\bibnamefont {Makhlin}},
	  \bibinfo {author} {\bibfnamefont {B.~N.}\ \bibnamefont {Narozhny}}, \bibinfo
	  {author} {\bibfnamefont {G.}~\bibnamefont {Sch{\"o}n}}, \ and\ \bibinfo
	  {author} {\bibfnamefont {A.}~\bibnamefont {Shnirman}},\ }\href {\doibase
	  10.1016/j.aop.2015.07.006} {\bibfield  {journal} {\bibinfo  {journal} {Annals
	  of Physics}\ }\textbf {\bibinfo {volume} {361}},\ \bibinfo {pages} {401}
	  (\bibinfo {year} {2015})}\BibitemShut {NoStop}%
	\end{thebibliography}

\end{document}